\documentclass[english]{article}
\usepackage[utf8]{inputenc}
\usepackage{fullpage} 
\usepackage{array}

\usepackage{bm}
\usepackage{amsmath}
\usepackage{amssymb}
\usepackage{graphicx}
\usepackage{caption}
\usepackage{subfigure}
\usepackage{tablefootnote}
\usepackage{setspace}
\usepackage[framed]{mcode}
\usepackage{xspace}
\usepackage{hyperref}
\usepackage{authblk}

\newcommand{\Rset}{\mathbb{R}}
\newcommand{\Nset}{\mathbb{N}}

\newcommand{\yy}{\mathbf{y}}
\newcommand{\ii}{\mathbf{i}}
\newcommand{\jj}{\mathbf{j}}
\newcommand{\ee}{\mathbf{e}}
\newcommand{\dd}{\, \mathrm{d}}
\newcommand{\pp}{\mathbf{p}}

\newcommand{\SGK}{\Ladd{SGMK}\xspace}

\newtheorem{example}{Example}
\newtheorem{remark}{Remark}

\newcommand{\Ladd}[1]{{#1}}

\title{The Sparse Grids Matlab Kit - a Matlab implementation of sparse grids for high-dimensional function approximation and uncertainty quantification}
\author[1,2]{Chiara Piazzola}
\author[1]{Lorenzo Tamellini}
\affil[1]{Istituto di Matematica Applicata e Tecnologie Informatiche ``E. Magenes'', Consiglio Nazionale delle Ricerche, Via Ferrata, 5/A 27100, Pavia, Italy}
\affil[2]{Department of Mathematics, Technical University of Munich, 
	Boltzmannstra\ss e, 3 85748, Garching bei M\"unchen, Germany 
	\protect \\
	\nolinkurl{chiara.piazzola@tum.de}, \nolinkurl{tamellini@imati.cnr.it}}
\date{}

\begin{document}
	
	\maketitle
	
\section*{Abstract}
 \Ladd{The Sparse Grids Matlab Kit provides a Matlab implementation of sparse grids, 
  and can be used for approximating high-dimensional functions and, in particular, for surrogate-model-based uncertainty quantification. 
  It is lightweight, high-level and easy to use, good for quick prototyping and teaching;
  however, it is equipped with some features that allow its use also in realistic applications.
  The goal of this paper is to provide an overview of the data structure and of the mathematical aspects
  forming the basis of the software, as well as comparing the current release of our package to similar available software}.

\section{Introduction}

\Ladd{The reliability of computer simulations in science and engineering
  crucially depends on having a precise knowledge of the input parameters of such simulations, such as coefficients, forcing terms, initial and boundary conditions,
  shape of the computational domain and so on. However, in practical scenarios it is quite common to have only a partial knowledge of said parameters,
  due e.g. to measurement errors, temporal/monetary/technological constraints on running an experimental campaign, or intrinsic randomness of certain quantities.
  It is therefore useful to model these uncertain parameters as random variables, and to think of a computer simulation as a function $f$ that
  associates the corresponding simulation outputs to each possible value of such uncertain input parameters.
  The values of $f$ (i.e., the simulation outputs) are then also random/uncertain quantities, whose statistical properties should be assessed.
  This kind of analysis is called Uncertainty Quantification (UQ).}

\Ladd{UQ analyses pose significant theoretical and computational challenges, mainly due to two facts: 1) the number of uncertain parameters could be very large,
  i.e., $f$ could be a high-dimensional function; 2) evaluating $f$ requires running a computer simulation, which could be computationally expensive.
  One major research direction that has been explored in literature to deal with these issues 
  consists in a two-step procedure: 1) deriving an approximation of $f$, say $\widehat{f}$ 
  (that can go under different names, each carrying different meaning nuances 
  also depending on the scientific community:
  surrogate model, proxy model, response function, reduced order model),
  that is ideally both cheap to obtain and much faster to evaluate than $f$;
  2) computing the statistical properties of $\widehat{f}$ instead of those of $f$.
  Among the several surrogate modeling techniques available in literature (see e.g. \cite{ghanem:UQbook,smith:UQbook}), 
  we focus here on the so-called sparse grid method.}

\Ladd{More specifically, the aim of this manuscript is to introduce the Sparse Grids Matlab Kit (\SGK)
  as a tool for high-dimensional function approximation and UQ.
  The \SGK is freely available under the BSD2 license on Github \cite{sgmk:github},
  and full resources (past and current releases, user manual \cite{piazzola.eal:user.manual},  
  and other release-related material including source code from selected publications that have used the \SGK)
  are available on a dedicated website \cite{sgmk:website};}
the first version was released in 2014 (14-4 ``Ritchie''), and the current version was released in 2023 
(23-5 ``Robert''). It is written in Matlab, and its compatibility with Octave has been tested;
it is extensively documented and commented (release 23-5 has about 9800 lines of code and 5300 lines of comment).
The release contains several tutorials and a testing unit is also available.

From a mathematical point of view, the package implements the combination technique form of sparse grids.
It is a high-level package with a syntax quite close to the mathematical description of sparse grids
which makes it (hopefully) easy to use and, therefore, suitable for quick prototyping and didactic purposes
(for example, it has been used to write the codes of the book \cite{MartinezFrutos2018}).
\Ladd{However, we will point out a number of functionalities
  that make the \SGK usable for realistic UQ problems, as well as for problems with hundreds of random variables:
  an interface with the Matlab Parallel Toolbox; a strategy to recycle evaluations of $f$ that might be
  already available from previous computations; a so-called buffering strategy that improves the work allocation when
  $f$ depends on a very large number of random variables; full compatibility with the UM-Bridge protocol \cite{Seelinger2023:umbridge}
  for HTTP communication with external software for evaluating $f$.
  These features have been used for example to obtain the results shown in 
  \cite{colombo.eal:discontinuities,piazzola.eal:ferry-paper,chiappetta.eal:additive,seelinger.eal:kubernetes,nobile.eal:adaptive-lognormal,ernst.eal:collocation-logn}.
}

\Ladd{Although being general enough to be used for manipulation of high-dimensional functions in many frameworks,
  the \SGK is geared towards UQ, as already mentioned.
  In particular, it provides the following UQ tools:
  collocation points for several random variables (uniform, normal, exponential, gamma, beta, triangular);
  computation of generalized Polynomial Chaos Expansions 
  \cite{xiu.karniadakis:wiener,ernst:approximability,sudret:sobol} for such random variables; 
  computation of Sobol sensitivity indices \cite{sobol2001,archer.saltelli.sobol:anova,feal:compgeo,sudret:sobol}; 
  approximation of gradients and Hessians of a function, which could be useful for a number of UQ tasks, such as local sensitivity analysis \cite{cacuci:sensitivity},
  detection of active subspaces \cite{constantine:book},
  Maximum-A-Posteriori estimation of uncertain parameters \cite{thanh-bui.gattas:MCMC,stuart:acta.bayesian,piazzola.eal:SIR,kim.eal:hippylib},
  and Markov-Chain Monte-Carlo (MCMC) sampling \cite{petra.ghattas:ice.sheet.MCMC,stuart:acta.bayesian,kim.eal:hippylib}.
  It is also straightforward to connect the \SGK functions with built-in Matlab functions for minimization and MCMC sampling,
  as well as for standard UQ tasks such as computation of histograms and approximation of probability density functions.
}

\Ladd{The \SGK} belongs to the same niche \Ladd{as} a number of other \Ladd{packages for surrogate modeling and UQ purposes};
we provide a (knowingly incomplete) list in \Ladd{Table}\ \ref{tab:software}.
The \Ladd{package} in the table closest to the \SGK (in terms of language, functionalities and usability) is probably Spinterp, which is however no longer maintained
and does not implement any UQ function. A deeper discussion is reported in \Ladd{Section} \ref{section:comparison}, where a closer comparison in terms of functionalities
\Ladd{is given} between the \SGK and the other Matlab-based sparse-grids/UQ software  \Ladd{in Table \ref{tab:software}}
(either natively implemented in Matlab or providing interfaces to software written in C++/Python),
i.e., SG++, Spinterp, Tasmanian.

\begin{table}[tb]
  \centering
  \begin{tabular}{l>{\raggedright}p{2.8cm}ll}
    \hline
    \textbf{Name} & \textbf{Language} 	& \textbf{Ref.} 		& \textbf{Webpage}\\
    \hline
    Dakota 	  & C++		      	& \cite{dakota} 		& \url{https://dakota.sandia.gov} \\
    PyApprox	  & Python	      	& \cite{jakeman:pyapprox}       & \url{https://pypi.org/project/pyapprox} \\	
    MUQ	          & C++, Python       	& \cite{MUQ} 			& \url{https://mituq.bitbucket.io}\\  
    UQLab         & Matlab            	& \cite{UQLab} 			& \url{https://uqlab.com} \\
    ChaosPy       & Python            	& \cite{ChaosPy,ChaosPy1}	& \url{https://chaospy.readthedocs.io}\\
    SG++          & Python, Matlab, Java, C++  	& \cite{SG++} 		& \url{https://sgpp.sparsegrids.org/} \\
    Spinterp      & Matlab            	& \cite{klimke:thesis,spinterp,spinterp:manual}	& \url{http://calgo.acm.org/847.zip}  \\
    UQTk          & C++, Python 	& \cite{UQTk,UQTk1}    		& \url{https://sandia.gov/uqtoolkit}\\ 
    Tasmanian     & C++, Python, Matlab, Fortran 90/95	   & \cite{tasmanian,tasmanianPaper1,tasmanianPaper2} & \url{https://github.com/ORNL/TASMANIAN} \\
    OpenTURNS     & C++, Python       & \cite{OpenTURNS}                & \url{https://openturns.github.io/www/index.html}\\
    URANIE        & C++, Python       & \cite{URANIE}                   & \url{https://www.salome-platform.org/?page_id=2019}\\
    UncertainSCI  & Python            & \cite{UncertainSCI}             & \url{https://www.sci.utah.edu/cibc-software/uncertainsci.html}\\
    \hline
  \end{tabular}
  \caption{List of high-dimensional approximation / UQ-related software.
  \Ladd{Note that the webpage reported for Spinterp corresponds to the last officially released version, to the best of knowledge of the authors of this manuscript.
        A later version can be found at \url{https://people.sc.fsu.edu/~jburkardt/m_src/spinterp/spinterp.html}}.}
  \label{tab:software}
\end{table}

The rest of the paper is organized as follows. \Ladd{Section}\ \ref{sect:sg} introduces the minimal mathematical background necessary
to understand the entities implemented in the \SGK. \Ladd{Section}\ \ref{sect:sg_data_structure} covers how sparse grids are generated in the \SGK
and \Ladd{the data structure used to store them. Note,} in particular, that the \SGK provides two mechanisms to generate sparse grids: a-priori and adaptive a-posteriori.
\Ladd{Section}\ \ref{sect:features} discusses the main functionalities available in the \SGK,
with special emphasis on \Ladd{evaluation} recycling, parallelization, interface with the UM-Bridge protocol and \Ladd{computation of} 
polynomial chaos expansions. \Ladd{Section}\ \ref{section:comparison} contains the 
comparison with the other Matlab sparse-grids/UQ software available in the literature.
\Ladd{Finally, Section \ref{section:conclusions} draws some conclusions.
  In general, we will keep the discussion on a high-level, language-independent tone. However,
  from time to time, we will give references to specific parts the user manual \cite{piazzola.eal:user.manual},
  where interested readers can find implementation details.}

\section{Mathematical basics of sparse grids}\label{sect:sg}
We consider the two problems of a) approximating and b) computing weighted integrals,
\Ladd{i.e., expected values and higher-order moments,} of (the components of) a function $ f: \Rset^N \rightarrow \Rset^V$
given some samples of $f$ whose location we are free to choose. More specifically, we assume that $f$ depends on $N$
random variables $\yy = (y_1, \ldots,y_N) \in \Gamma$, with $\Gamma = \Gamma_1 \times \ldots \times \Gamma_N \subset \Rset^N$
being the set of all possible values of $\yy$. We denote by $\rho_n:\Gamma_n \rightarrow \Rset^+$
the probability density function (pdf) of each variable $y_n$, $n=1,\ldots,N$
and assume independence of $y_1,\ldots,y_N$, such that the joint pdf of $\yy$ is $\rho(\yy) = \prod_{n=1}^N \rho_n(y_n), \forall \yy \in \Gamma$.
\Ladd{Note that the assumption of independence is kept for simplicity but is actually not needed.
  Indeed, approximations of $f$ can be built based only on the marginal probability density functions of $y_1, \ldots,y_N$, 
  and, while independence is needed for quadrature, in case $y_1, \ldots,y_N$ are correlated it is possible to use the theory of copulas \cite{nelsen:copulas}
  to introduce a change of variables where the new random variables are independent uniform random variables, and to set the sparse grid in this new set of variables.}

\Ladd{The first step in building a sparse grid is to define a set of collocation knots for each variable $y_n$.
  We thus introduce the univariate discretization level $i_n \in \Nset_+$ and a non-decreasing function, called ``level-to-knots function'',
  that specifies the number of collocation knots to be used for each random variable at discretization level $i_n$, i.e.,}
\begin{equation}\label{eq:lev2knots}
	m:\Nset_+ \rightarrow \Nset_+, \ i_n \mapsto m(i_n). 
\end{equation}  
Then, we denote by $\mathcal{T}_{n,i_n}$ the set of $m(i_n)$ knots along $y_n$, i.e.,
\begin{equation}\label{eq:1d-nodes}
\mathcal{T}_{n,i_n} = \left\{y_{n,m(i_n)}^{(j_n)}: j_n=1, \ldots, m(i_n)\right\} \quad \text{ for } n=1,\ldots,N.  
\end{equation}
Typical examples of level-to-knots functions are:
  \begin{align}
    m(i)=i  			& \quad \text{(\emph{linear})}, \label{eq:m-linear}\\
    m(i)=2(i-1)+1  		& \quad \text{(\emph{$2$-step})}, \label{eq:m-twostep}\\
    m(1)=1, \, m(i) = 2^{i-1}+1 \text{ for } i>1 	& \quad \text{(\emph{doubling})}. \label{eq:m-doubling}
  \end{align}
The knots of $\mathcal{T}_{n,i_n}$ are usually chosen according to the probability distribution of the random variables $\rho_n$,
\Ladd{to guarantee a good convergence rate of the resulting sparse grid approximation, see for example
  \cite{nobile.tempone.eal:sparse,nobile.tempone.eal:aniso,ernst.eal:collocation-logn}.}
For efficiency reasons, it is beneficial if the sequences of knots are nested,
i.e., if $\mathcal{T}_{n,i_n} \subset \mathcal{T}_{n,j_n}$ with $j_n\geq i_n$. However, the \SGK does not require
\Ladd{these sequences to be nested,} contrary to other software \Ladd{(see Section \ref{section:comparison} for details).}

Next, we introduce $N$-dimensional tensor grids obtained by taking the Cartesian product of the $N$ univariate sets of knots just introduced.
For this purpose we collect the discretization levels $i_n$ in a multi-index
$\ii = [i_1,\ldots,i_N] \in \mathbb{N}^N_+$ and denote the corresponding tensor grid by $\mathcal{T}_{\ii} = \bigotimes_{n=1}^{N} \mathcal{T}_{n,i_n}$.
Using standard multi-index notation, we can then write 
\[
  \mathcal{T}_{\ii} = \left\{\yy_{m(\ii)}^{(\jj)}\right\}_{\jj \leq m(\ii)},
  \quad  \text{ with } \quad
  \yy_{m(\ii)}^{(\jj)} = \left[y_{1,m(i_1)}^{(j_1)}, \ldots, y_{N,m(i_N)}^{(j_N)}\right]
  \text{ and } \jj \in \mathbb{N}^N_+,
\]
where $m(\ii) = \left[m(i_1),\,m(i_2),\ldots,m(i_N) \right]$ and $\jj \leq m(\ii)$ means that $j_n \leq m(i_n)$ for every $n = 1,\ldots,N$.

A tensor grid approximation of $f(\yy)$ based on global Lagrangian interpolants collocated at these grid knots
can then be written in the following form  
\begin{equation} \label{eq:interp_tensor}
\mathcal{U}_{\ii}(\yy) := \sum_{\jj \leq m(\ii)} f\left(\yy_{m(\ii)}^{(\jj)}\right) \mathcal{L}_{m(\ii)}^{(\jj)}(\yy),
\end{equation}
where $\left\{ \mathcal{L}_{m(\ii)}^{(\jj)}(\yy) \right\}_{\jj \leq m(\ii)}$ are $N$-variate Lagrange basis polynomials, 
defined as tensor products of univariate Lagrange polynomials, i.e.,  
\begin{equation}\label{eq:lagrange_pol}
  \mathcal{L}_{m(\ii)}^{(\jj)}(\yy) = \prod_{n=1}^{N} l_{n,m(i_n)}^{(j_n)}(y_n)
  \quad \text{ with } \quad
  l_{n,m(i_n)}^{(j_n)}(y_n) = \prod_{k=1, k\neq j_n}^{m(i_n)} \frac{y_n-y_{n,m(i_n)}^{(k)}}{y_{n,m(i_n)}^{(j_n)}-y_{n,m(i_n)}^{(k)}}. 
\end{equation}
\Ladd{Note that other choices could be made here, namely replacing Lagrange polynomials in \eqref{eq:interp_tensor}
by, \Ladd{for example,} splines as in \cite{rehme.eal:splines}, or trigonometric polynomials as in \cite{stoyanov:adaptive2};
we discuss this issue further in Section \ref{section:comparison} when comparing different sparse grid software.}

Similarly, the tensor grid quadrature of $f(\yy)$, i.e., the approximation of its weighted integral \Ladd{(expected value)},
can be computed by taking the integral of the Lagrangian interpolant in \eqref{eq:interp_tensor}:
\begin{align}\label{eq:quad_tensor}
\mathcal{Q}_{\ii} := \int_{\Gamma} \mathcal{U}_{\ii}(\yy) \rho(\yy) \dd \yy 
& = \sum_{\jj \leq m(\ii)} f\left(\yy_{m(\ii)}^{(\jj)}\right) \left( \prod_{n=1}^{N} \int_{\Gamma_n} l_{n,m(i_n)}^{(j_n)}(y_n) \rho(y_n) \dd y_n\right) \nonumber \\
& = \sum_{\jj \leq m(\ii)} f\left(\yy_{m(\ii)}^{(\jj)}\right)  \left( \prod_{i=1}^{N} \omega_{n,m(i_n)}^{(j_n)} \right)
= \sum_{\jj \leq m(\ii)} f\left(\yy_{m(\ii)}^{(\jj)}\right) \omega_{m(\ii)}^{(\jj)},
\end{align}
where $\omega_{n,m(i_n)}^{(j_n)}$ are the standard quadrature weights obtained by computing the integrals of the associated univariate
Lagrange polynomials, and $\omega_{m(\ii)}^{(\jj)}$ are their multivariate counterparts. 

Naturally, the approximations $\mathcal{U}_{\ii}$ and $\mathcal{Q}_{\ii}$ are more and more accurate the higher the number of 
collocation knots in each random variable and, therefore, one would ideally choose all the components of $\ii$ to be large,
say $\ii = \ii^\star$ with $i^\star_n \gg 1, \forall n=1,\ldots,N$.
\Ladd{However, obtaining these approximations could be too compuationally expensive} even for $N$ moderately large,
due to fact that they would require $\prod_{n=1}^N m(i_n^\star)$ evaluations of $f$, i.e., their cost grows exponentially in $N$.

To circumvent this issue, the sparse grid method replaces $\mathcal{U}_{\ii^\star}$ with a linear combination of multiple coarser $\mathcal{U}_{\ii}$,
and similarly for $\mathcal{Q}_{\ii^\star}$ (from now on we use the generic symbol $\mathcal{F}_{\ii}$ to denote both $\mathcal{U}_{\ii}$ and $\mathcal{Q}_{\ii}$).
To this aim we introduce the so-called ``detail operators'' (univariate and multivariate). 
They are defined as follows, with the understanding that $\mathcal{F}_{\ii} (\yy) = 0$ when at least one component of $\ii$ is zero.
Thus, \Ladd{let $\ee_n$} the $n$-th canonical multi-index, i.e., $(\ee_n)_k = 1$ if $n=k$ and 0 otherwise, and define
\begin{alignat}{2}
&\text{\textbf{Univariate detail:}}
&& \Delta_n[\mathcal{F}_{\ii}]=\mathcal{F}_{\ii}-\mathcal{F}_{\ii-\ee_n}, \text{ with } 1 \leq n \leq N, \nonumber \\
&\text{\textbf{Multivariate detail: }}
&& \bm{\Delta}[\mathcal{F}_{\ii}] = \bigotimes_{n=1}^N \Delta_n[\mathcal{F}_{\ii}], \label{eq:details} 
\end{alignat}
where taking tensor products of univariate details amounts to composing their actions, i.e.,
\[
\bm{\Delta}[\mathcal{F}_{\ii}]
= \bigotimes_{n=1}^N \Delta_n[\mathcal{F}_{\ii}]
= \Delta_1\left[ \, \cdots \left[ \Delta_N\left[ \mathcal{F}_{\ii} \right] \, \right] \, \right].
\]
By replacing the univariate details with their definitions, we can then see that this implies
that the multivariate operators can be evaluated by evaluating certain full-tensor approximations  
$\mathcal{F}_{\ii}$, and then taking linear combinations:
\begin{align*}
\bm{\Delta}[\mathcal{F}_{\ii}]
& = \Delta_1\left[ \, \cdots \left[ \Delta_N\left[ \mathcal{F}_{\ii} \right] \, \right] \, \right] 
= \sum_{\jj \in \{0,1\}^N} (-1)^{\lVert\jj\rVert_1} \mathcal{F}_{\ii-\jj}.
\end{align*}
Observe that by introducing these detail operators a hierarchical decomposition of $\mathcal{F}_{\ii}$ can be obtained;
indeed, the following telescopic identity holds true:
\begin{equation}\label{eq:telescopic_sum}
\mathcal{F}_{\ii} = \sum_{\jj \leq \ii} \bm{\Delta}[\mathcal{F}_{\jj}].
\end{equation}

\begin{example}[Telescopic identity]\label{example:tensor-grid}
  As an example of\, \Ladd{\eqref{eq:telescopic_sum},} 
  consider the case $N=2$.
Recalling that by definition $\mathcal{F}_{[j_1,j_2]} = 0$ when either $j_1=0$ or $j_2=0$, it can be seen that
\begin{align*}
\sum_{[j_1,j_2] \leq [2,2]} \bm{\Delta}[\mathcal{F}_{[j_1,j_2]}] 
= & \bm{\Delta}[\mathcal{F}_{[1,1]}]
+ \bm{\Delta}[\mathcal{F}_{[1,2]}]
+ \bm{\Delta}[\mathcal{F}_{[2,1]}]
+ \bm{\Delta}[\mathcal{F}_{[2,2]}] \\
= & \mathcal{F}_{[1,1]}
+ ( \mathcal{F}_{[1,2]} - \mathcal{F}_{[1,1]} )
+ ( \mathcal{F}_{[2,1]} - \mathcal{F}_{[1,1]} ) 
+ (\mathcal{F}_{[2,2]} - \mathcal{F}_{[2,1]} - \mathcal{F}_{[1,2]} + \mathcal{F}_{[1,1]}) \nonumber \\
= & \mathcal{F}_{[2,2]}. \nonumber
\end{align*}
\end{example}

The crucial observation that allows \Ladd{us to obtain a sparse grid} is that, under suitable regularity assumptions for $f(\yy)$,
not all of the details in the hierarchical decomposition in \eqref{eq:telescopic_sum} contribute equally to the approximation,
i.e., some of them can be discarded and the resulting formula will retain good approximation properties at a fraction of the computational cost:
roughly speaking, the multi-indices to be discarded are those corresponding to ``high-order'' details, i.e., those for which $\| \jj \|_1 $
is sufficiently large, see, \Ladd{for example, \cite{b.griebel:acta}.
A simple way to discard high-order details would be for instance 
to replace the summation set $\{\jj \leq \ii\}$ in \eqref{eq:telescopic_sum} with $\{\| \jj \|_1 \leq  w\}$ for $w \in \Nset_+$ small enough,
as shown in the next Example \ref{example:sparse-grid}.}
\begin{example}[\Ladd{Discarding} high-order details]\label{example:sparse-grid}
  Following \Ladd{Example \ref{example:tensor-grid}} and replacing the constraint $[j_1,j_2] \leq [2,2]$ with $\| \jj \|_1 \leq 3$,
  we obtain the following approximation:
  \begin{align*}
    \mathcal{F}_{[2,2]} \approx \sum_{j_1+j_2 \leq 3} \bm{\Delta}[\mathcal{F}_{[j_1,j_2]}] 
    = \bm{\Delta}[\mathcal{F}_{[1,1]}]
    + \bm{\Delta}[\mathcal{F}_{[1,2]}]
    + \bm{\Delta}[\mathcal{F}_{[2,1]}]
    =  - \mathcal{F}_{[1,1]} + \mathcal{F}_{[1,2]} + \mathcal{F}_{[2,1]}.
  \end{align*} 
\end{example}
In general, upon collecting the multi-indices to be retained in the sum in a multi-index set $\mathcal{I} \subset \mathbb{N}_+^{N}$
the sparse grids approximation of $f$ and of its weighted integral can finally be written as (see, \Ladd{for example,} \cite{wasi.wozniak:cost.bounds}):
\begin{align}
  f(\yy) & \approx \mathcal{U}_{\mathcal{I}} (\yy)
           = \sum_{\ii \in \mathcal{I}} \bm{\Delta}[\mathcal{U}_{\ii}(\yy)]
           = \sum_{\ii \in \mathcal{I}} c_{\ii} \, \mathcal{U}_{\ii}\,, \quad c_{\ii}: = \sum_{\substack{\jj \in \{0,1\}^N \\ \ii+\jj \in \mathcal{I}}} (-1)^{\lVert\jj\rVert_1} \label{eq:sg_interp} \\
  \int_\Gamma f(\yy) \rho(\yy) \dd \yy
         & \approx \mathcal{Q}_{\mathcal{I}} (\yy)
           = \sum_{\ii \in \mathcal{I}} \bm{\Delta}[\mathcal{Q}_{\ii}(\yy)]
           = \sum_{\ii \in \mathcal{I}} c_{\ii} \mathcal{Q}_{\ii},  \label{eq:sg_quad}
\end{align}
and the sparse grid is defined as
\begin{equation}\label{eq:sg_grid} 
\mathcal{T}_\mathcal{I} =  \bigcup_{\substack{\ii \in \mathcal{I} \\ c_{\ii} \neq 0 }} \mathcal{T}_{\ii}.
\end{equation}

\begin{remark}\label{rem:interpolation}
  The sparse grid approximation in \eqref{eq:sg_interp} is not necessarily an interpolant operator, i.e.,
  \Ladd{the equality $f(\yy_k) = \mathcal{U}_{\mathcal{I}} (\yy_k)$ for $\yy_k \in \mathcal{T}_\mathcal{I}$ does not necessarily hold true.}
  More specifically, a sparse grid is interpolatory only if it is built with nested knots.
  In the following, with a slight abuse of terminology we will nonetheless refer to the operation of evaluating
  $\mathcal{U}_{\mathcal{I}} (\yy)$ by means of \eqref{eq:sg_interp} as sparse grid interpolation.
  Another commonly used terminology, especially in the field of uncertainty quantification,
  is to refer to $\mathcal{U}_{\mathcal{I}} (\yy)$ as \Ladd{the} sparse grid surrogate model \Ladd{of $f$}.
\end{remark}

\begin{figure}
 	\centering
 	\includegraphics[width = 0.25\textwidth]{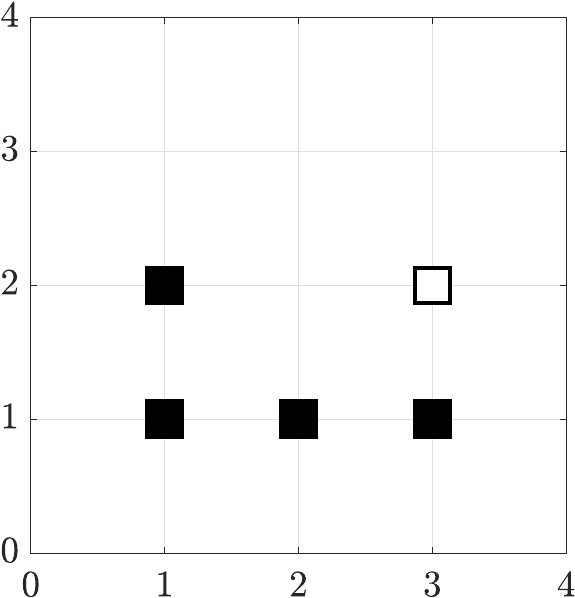}
 	\caption{{D}ownward closedness of a multi-index set. The set of the multi-indices marked in \Ladd{black} is downward closed.
          Instead, the multi-index $[3,2]$ (in \Ladd{white}) violates the rule in \eqref{eq:downward_closed}:
          the multi-index $[2,2] = [3,2] - \ee_1$ is not contained in the multi-index set and hence the set of blue and red multi-indices is not downward closed. }
 	\label{fig:downward_closed_set}
 \end{figure}

The right-most equalities in \eqref{eq:sg_interp} and \eqref{eq:sg_quad} are known as the combination technique form
of the sparse grids approximation and quadrature (see \cite{Griebel.schneider.zenger:combination}), which is the form implemented in the \SGK.
Another possibility would be to implement the first form, i.e., the sum of detail operators
$\sum_{\ii \in \mathcal{I}} \bm{\Delta}[\mathcal{U}_{\ii}(\yy)]$, which is known as the hierarchical form of sparse grids.
In particular, implementing the hierarchical form requires introducing a basis for the detail operators $\Delta[\mathcal{F}_{\ii}]$
in \eqref{eq:details} rather than for the tensor interpolants $\mathcal{U}_{\ii}$:
this request naturally suggests using the hierarchical form of piecewise polynomials
\Ladd{as a basis (for example, the classical hat-functions). In turn, this} opens the way to the so-called local adaptivity of sparse grids,
\Ladd{as discussed in} \cite{pflueger:adaptive,Eftekhari:locally.adapt,obersteiner:locally.adapt,ma.zabaras:locally.adapt}. 
Using piecewise polynomials to introduce a basis for the detail operators $\Delta[\mathcal{F}_{\ii}]$ is not mandatory though
and it would be possible to use global Lagrange polynomials even in this context, using the hierarchical form of Lagrange polynomials
described, \Ladd{for example,} in \cite{eigel.eal:convergence,chkifa:adaptive-interp}.
Note that the equivalence between the two forms is true only if $\mathcal{I}$ is chosen as downward closed, i.e.,
\begin{equation*}\label{eq:downward_closed}
\forall \mathbf{k} \in \mathcal{I}, \quad \mathbf{k} - \ee_n \in \mathcal{I} \text{ for every } n=1,\ldots,N \text{ such that } k_n > 1, 
\end{equation*}
see \Ladd{Figure} \ref{fig:downward_closed_set}. 
The choice of implementing the combination technique instead of the hierarchical form
allows to keep the data structure to a minimum, and guarantees ease of use and high-level, ``close-to-the-math'' coding.
\Ladd{In Section \ref{section:comparison} we examine a number of existing software packages and specify
  whether they implement the combination technique or the hierarchical form of sparse grids (or both).}

Coming back to the choice of the set $\mathcal{I}$, the optimal choice depends on the decay rate of the \Ladd{norm of the} detail operators, which in turn depends
on the regularity of $f$, see, \Ladd{for example,} \cite{b.griebel:acta,nobile.eal:optimal-sparse-grids,chkifa:adaptive-interp}.
One classical choice of \Ladd{the} set $\mathcal{I}$ is the following one:
  \begin{equation}
    \label{eq:TD_set_aka_smolyak}
    \mathcal{I}_{\text{sum}}(w) = \left\{ \ii \in \Nset^N_+ : \sum_{n=1}^N (i_n-1) \leq w  \right\}  ,  
  \end{equation}
  for some $w \in \Nset$.
In particular, together with the doubling level-to-knots, \eqref{eq:m-doubling}, this results in the famous Smolyak grid.
\Ladd{Examples of other choices are:}
  \begin{align}
    \text{Tensor product set: }\, & \mathcal{I} = \left\{ \ii \in \Nset^N_+ : \max_{n=1}^N g_n(i_n-1) \leq w  \right\},  \label{eq:TP-set}\\
    \text{Total degree set: }\,	 & \mathcal{I} = \left\{ \ii \in \Nset^N_+ : \sum_{n=1}^N g_n(i_n-1) \leq w  \right\},  \label{eq:TD-set} \\
    \text{Hyperbolic cross set: }\, & \mathcal{I} = \left\{ \ii \in \Nset^N_+ : \prod_{n=1}^N (i_n)^{g_n} \leq w  \right\}, \label{eq:HC-set} \\
     \text{Multi-index box set: }\, & \mathcal{I} = \left\{ \ii \in \Nset^N_+ : i_n \leq b_n  \right\}, \label{eq:box-set}
  \end{align}
  for $g_1,\ldots,g_N \in \mathbb{R}$, $w,b_1\ldots,b_N \in \mathbb{N}$,
  where, in particular, $g_1,\ldots,g_N$ are typically called anisotropy weights and $w$ sparse-grid level.
  For a thorough discussion \Ladd{of} the motivations for introducing the sets above, \Ladd{see, for example, \cite{back.nobile.eal:comparison}.}
  The set $\mathcal{I}$ could also be built adaptively based on suitable profit indicators; this leads to adaptive sparse grids algorithms, that will be discussed in
  \Ladd{Section} \ref{sect:sparse_grid_generation}.

 \begin{figure}
 	\centering
 	\subfigure[$\mathcal{T}_{[1,1]}$]{\includegraphics[width=0.2\textwidth]{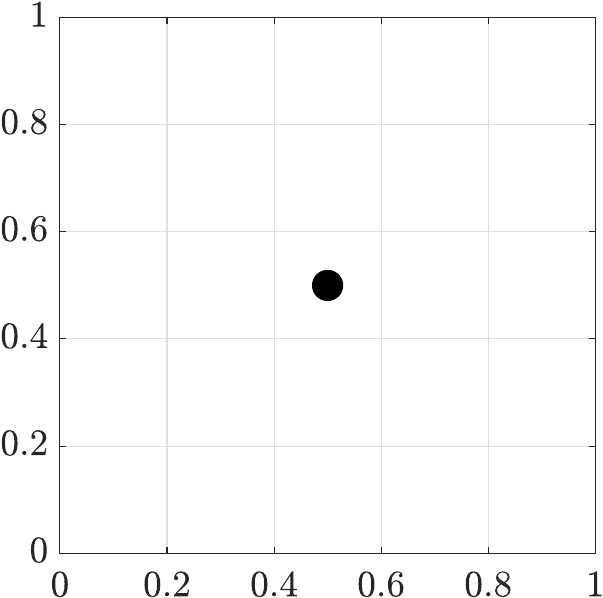}}\hspace{1.5em}
 	\subfigure[$\mathcal{T}_{[1,2]}$]{\includegraphics[width=0.2\textwidth]{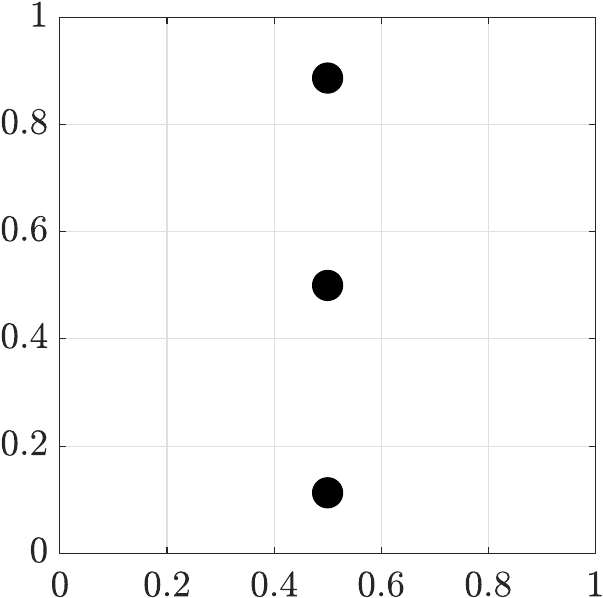}}\hspace{1.5em}
 	\subfigure[$\mathcal{T}_{[3,1]}$]{\includegraphics[width=0.2\textwidth]{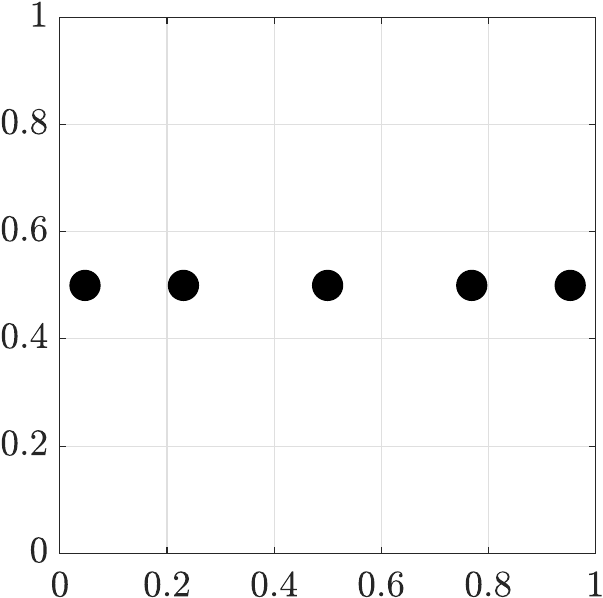}}\hspace{1.5em}
 	\subfigure[$\mathcal{T}_{\mathcal{I}}$]{\includegraphics[width=0.2\textwidth]{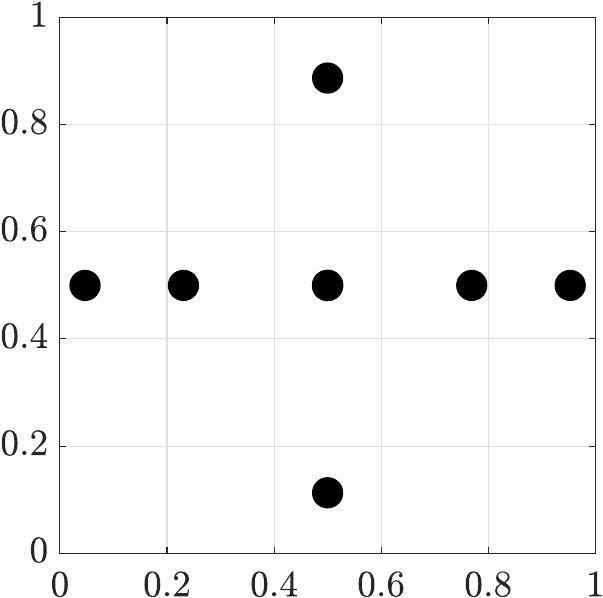}}\\
 	\subfigure[$\mathcal{T}_{[1,1]}$]{\includegraphics[width=0.2\textwidth]{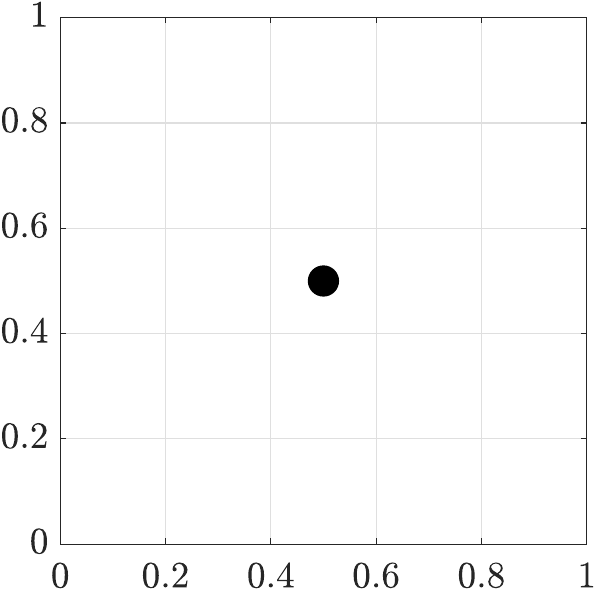}}\hspace{1.5em}
 	\subfigure[$\mathcal{T}_{[1,2]}$]{\includegraphics[width=0.2\textwidth]{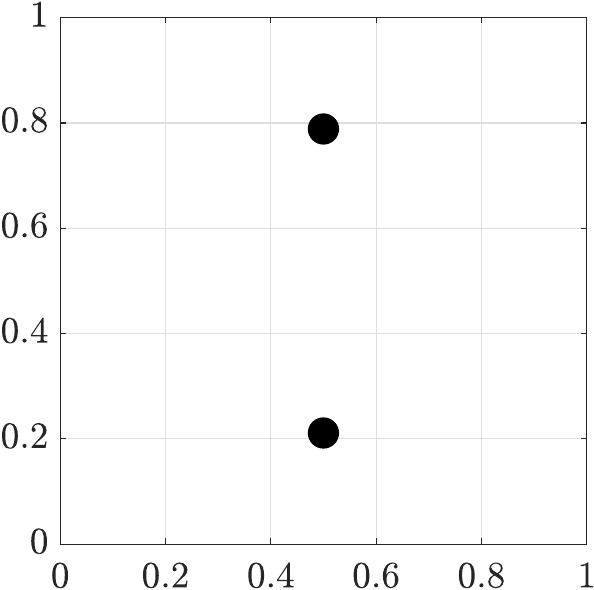}}\hspace{1.5em}
 	\subfigure[$\mathcal{T}_{[3,1]}$]{\includegraphics[width=0.2\textwidth]{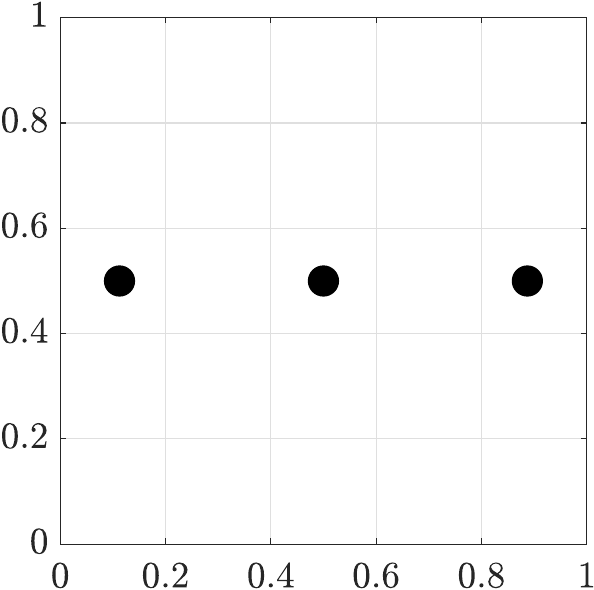}}\hspace{1.5em}
 	\subfigure[$\mathcal{T}_{\mathcal{I}}$]{\includegraphics[width=0.2\textwidth]{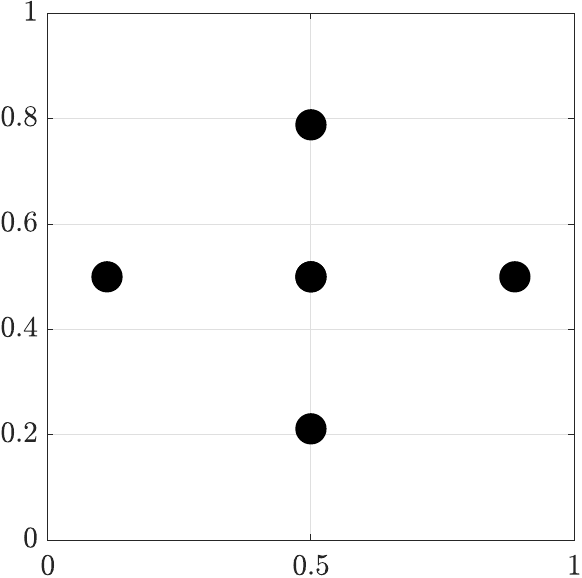}}        
 	\caption{Top row: tensor grids (panels a,b,c) and the sparse grid (panel d) for Example \ref{example:set-and-combitec} with
            level-to-knots doubling. Bottom row: tensor grids (panels e,f,g) and the sparse grid (panel h) for the same Example with
            level-to-knots linear.}
 	\label{fig:SG_comb}
 \end{figure}

\begin{example}\label{example:set-and-combitec}
Let\, $N=2$ and consider the downward closed multi-index set reported in \Ladd{Figure} \ref{fig:downward_closed_set}, 
i.e., $\mathcal{I} = \{[1,1],[1,2],[2,1],[3,1]\}$. \Ladd{We first show} the combination technique form of the sparse grid approximation
and quadrature in \eqref{eq:sg_interp} and \eqref{eq:sg_quad}, respectively.  
 We use again the generic symbol $\mathcal{F}_{\ii}$ to denote both $\mathcal{U}_{\ii}$ and $\mathcal{Q}_{\ii}$ and obtain 
 \[
 \mathcal{F}_{\mathcal{I}} (\yy) = c_{[1,1]}\mathcal{F}_{[1,1]}(\yy) + c_{[1,2]}\mathcal{F}_{[1,2]}(\yy) + c_{[2,1]} \mathcal{F}_{[2,1]}(\yy) + c_{[3,1]} \mathcal{F}_{[3,1]}(\yy) 
 \]
 with 
 \begin{align*}
 c_{[1,1]} & = (-1)^{\lVert [0,0] \rVert_1} + (-1)^{\lVert [1,0] \rVert_1} + (-1)^{\lVert [0,1] \rVert_1} = -1, \\
 c_{[1,2]} & = (-1)^{\lVert [0,0] \rVert_1} = +1, \\
 c_{[2,1]} & = (-1)^{\lVert [0,0] \rVert_1} + (-1)^{\lVert [1,0] \rVert_1} = 0, \\
 c_{[3,1]} & = (-1)^{\lVert [0,0] \rVert_1} = +1. 
 \end{align*}
 Since $c_{[2,1]}=0$, only three Lagrangian interpolant/quadrature operators explicitly
 appear in the combination technique formulas \eqref{eq:sg_interp} and \eqref{eq:sg_quad}, i.e.,
 \[
 \mathcal{F}_{\mathcal{I}} (\yy) = - \mathcal{F}_{[1,1]}(\yy) + \mathcal{F}_{[1,2]}(\yy) + \mathcal{F}_{[3,1]}(\yy),
 \]
 and only the corresponding three tensor grids contribute to the sparse grid (cf. \eqref{eq:sg_grid}).
 Then, considering the doubling level-to-knots function \eqref{eq:m-doubling},
 the resulting tensor grids consist of one, three, and five grid knots, respectively: 
 \begin{align*}
 \mathcal{T}_{[1,1]} & = \left\{ \left[y_{1,1}^{1} \ y_{2,1}^{1}\right] \right\} ,\\
 \mathcal{T}_{[1,2]} & = \left\{ \left[y_{1,1}^{1} \ y_{2,3}^{1}\right],\ \left[y_{1,1}^{1} \ y_{2,3}^{2}\right],\ \left[y_{1,1}^{1} \ y_{2,3}^{3}\right] \right\},\\
 \mathcal{T}_{[3,1]} & = \left\{ \left[y_{1,5}^{1} \ y_{2,1}^{1}\right],\ \left[y_{1,5}^{2} \ y_{2,1}^{1}\right],\ \left[y_{1,5}^{3} \ y_{2,1}^{1}\right],\ \left[y_{1,5}^{4} \ y_{2,1}^{1}\right],\ \left[y_{1,5}^{5} \ y_{2,1}^{1}\right] \right\}. 
 \end{align*}
 Choosing Gauss-Legendre knots as univariate collocation knots on $\Gamma_1 = \Gamma_2 = [0,1]$
 leads to the tensor grids displayed in \Ladd{Figure} \ref{fig:SG_comb}a,b,c and the sparse grid of \Ladd{Figure} \ref{fig:SG_comb}d.
 If instead we consider the linear level-to-knots function \eqref{eq:m-linear}, the three tensor grids will
 consist of one, two and three knots
 \begin{align*}
 \mathcal{T}_{[1,1]} & = \left\{ \left[y_{1,1}^{1} \ y_{2,1}^{1}\right] \right\} ,\\
 \mathcal{T}_{[1,2]} & = \left\{ \left[y_{1,1}^{1} \ y_{2,2}^{1}\right],\ \left[y_{1,1}^{1} \ y_{2,2}^{2}\right] \right\},\\
 \mathcal{T}_{[3,1]} & = \left\{ \left[y_{1,3}^{1} \ y_{2,1}^{1}\right],\ \left[y_{1,3}^{2} \ y_{2,1}^{1}\right],\ \left[y_{1,3}^{3} \ y_{2,1}^{1}\right] \right\}. 
 \end{align*}
 and the same choice of Gauss-Legendre knots as univariate collocation knots on $\Gamma_1 = \Gamma_2 = [0,1]$
 leads to the tensor grids displayed in \Ladd{Figure}\ \ref{fig:SG_comb}e,f,g and the sparse grid of \Ladd{Figure} \ref{fig:SG_comb}h.
\end{example}

\section{The Sparse Grids Matlab Kit: sparse grid data structure}\label{sect:sg_data_structure}

As described in the previous section, defining a sparse grid requires choosing a family $\mathcal{T}_{n,i_n}$ of collocation knots for each
variable $y_n$, a level-to-knots function $m(\cdot)$ and a multi-index set $\mathcal{I}$
to be input in the combination technique formulas of \eqref{eq:sg_interp} and \eqref{eq:sg_quad}. 
In the \SGK, the same steps are to be followed to define a sparse grid,
as can be seen in \Ladd{the \SGK user manual, Listing 1 in Section 3 and the subsequent discussion.} 
As already discussed, the knots should be chosen according to the distribution of each random variable $y_n$.
The \SGK supports uniform, normal, exponential, gamma, beta
and triangular  distributions, and provides at least two choices of collocation knots 
(namely, Gaussian and weighted Leja knots, see \cite{quarteroni.sacco.eal:numerical,narayan:Leja} respectively) \Ladd{for almost all of them};
more options are provided for specific choices of random variables, such as midpoint, equispaced and
Clenshaw--Curtis knots \cite{trefethen:comparison} for uniform random variables,
and Genz--Keister \cite{genz.keister:kpnquad} for normal distributions;
\Ladd{the user manual reports some discussion on the algorithms used to compute
the knots and quadrature weights for the various choices in Section 3.1.} 
Several choices are also available for the level-to-knots functions,
including but not limited to the ones reported in \eqref{eq:m-linear}-\eqref{eq:m-doubling},
as well as for generating the multi-index sets in \eqref{eq:TP-set}-\eqref{eq:box-set}.
\Ladd{We refer to Section 3 of the user manual for a thorough discussion on the various options available.} 

\begin{figure}[tbp]
	\begin{minipage}[t]{0.3\textwidth}
		\vspace{0pt}
		\centering
		\includegraphics[width=\textwidth,clip,trim = 120 325 350 50]{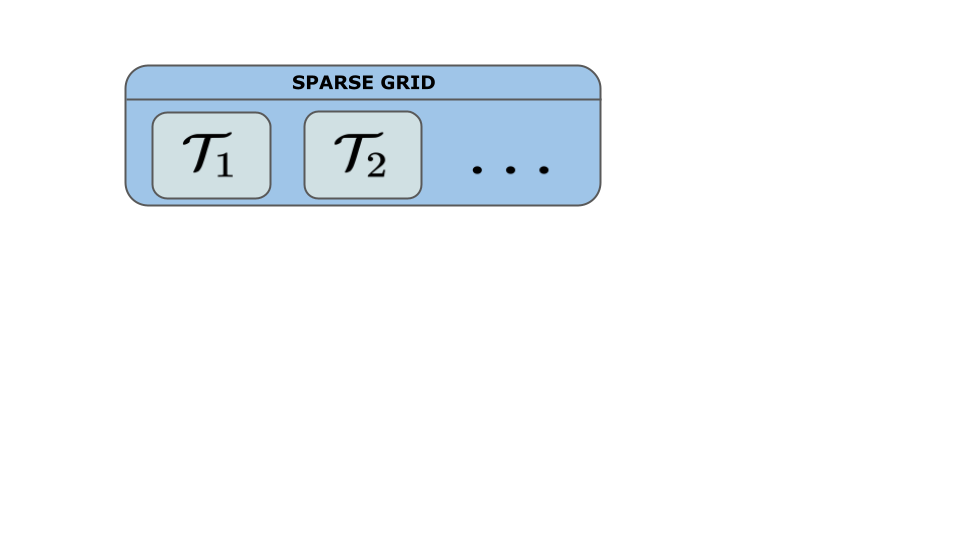}
		\captionof*{figure}{(a) Sparse grid structure array}
	\end{minipage}
	\begin{minipage}[t]{0.3\textwidth}
		\vspace{0pt}
		\centering
		\includegraphics[width=0.8\textwidth,clip,trim = 300 0 300 110]{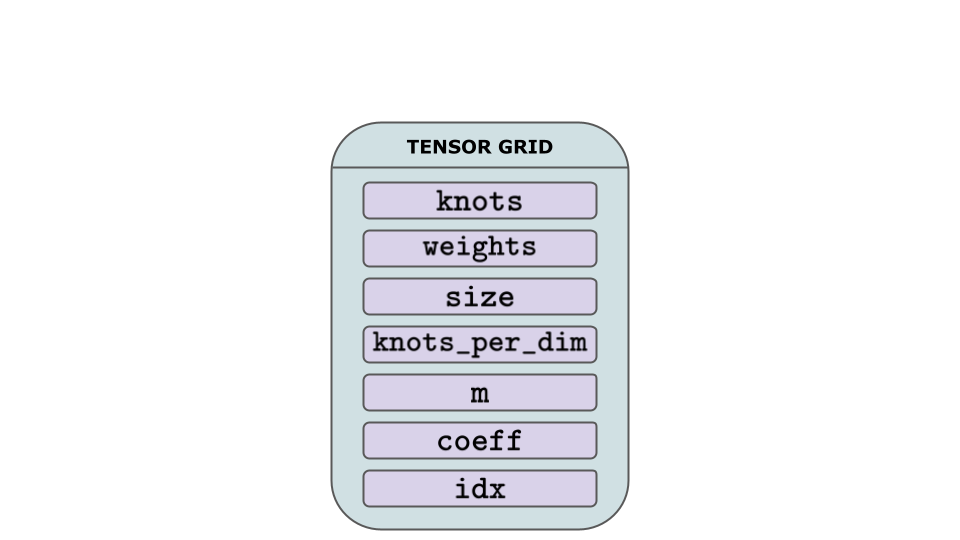}
		\captionof*{figure}{(b) Tensor grid structure}
	\end{minipage}
	\begin{minipage}[t]{0.3\textwidth}
		\vspace{0pt}
		\centering
		\includegraphics[width=0.8\textwidth,clip,trim = 300 80 300 110]{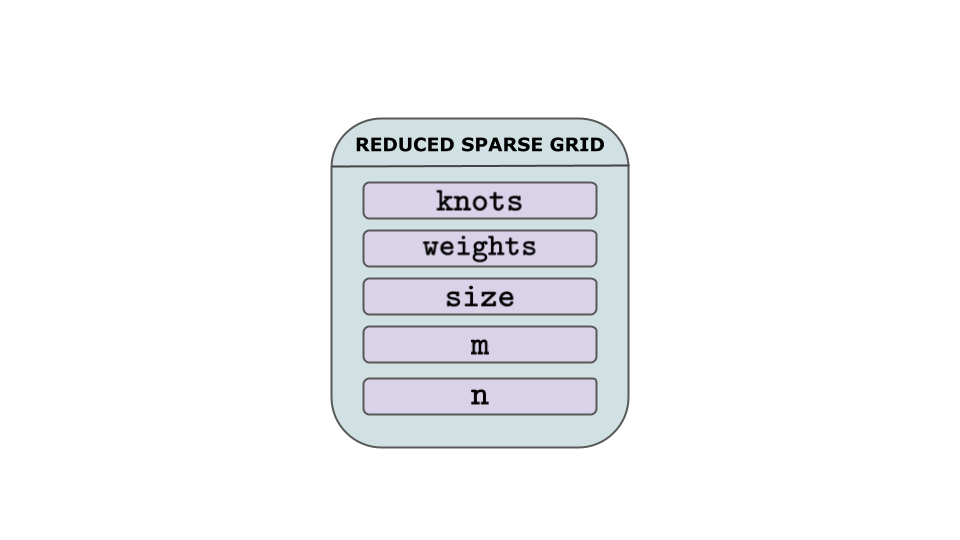}
		\captionof*{figure}{(c) Reduced sparse grid structure}
	\end{minipage}
	\caption{Sparse grid data structure: a sparse grid is stored \Ladd{in extended format} in a structure array \Ladd{(panel a)},
          each structure corresponding to one tensor grid \Ladd{(see panel b) that contains} 
          seven fields that identify the current grid. \Ladd{The data structure for the reduced format is instead shown in panel c.}}
	\label{fig:sg_data_structure}
\end{figure}

The data structure chosen to store \Ladd{a sparse grid} is also quite close to the mathematical formalism:
following the definition of a sparse grid in \eqref{eq:sg_grid} as a union of tensor grids, the \SGK stores a sparse grid
as a \Ladd{structure array}, where each \Ladd{component of the array} stores a single tensor grid
(see \Ladd{Figure} \ref{fig:sg_data_structure}a). Of course, only tensor grids whose coefficient
in the combination technique formula is non-zero \Ladd{are} stored,
see \eqref{eq:sg_interp} and \eqref{eq:sg_quad}.

The \Ladd{structures} storing a tensor grid are also quite minimal and contain only seven fields.
\Ladd{Following the ordering in Figure\ \ref{fig:sg_data_structure}b, the fields of the structure are:}
a matrix storing the knots of the tensor grid $\mathcal{T}_{\ii}$ \Ladd{(each knot being a column)},
a vector \Ladd{collecting} the corresponding quadrature weights multiplied by the combination technique coefficient of the current tensor grid,
i.e., $\omega_{m(\ii)} c_{\ii}$, an integer \Ladd{recording} the number of knots/weights,
a cell array containing the the univariate sets of collocation knots $\mathcal{T}_{n,i_n}$,
\Ladd{a row vector containing} the number of knots in each dimension
an integer storing the coefficient $c_{\ii}$ of the combination technique formula, see \eqref{eq:sg_interp},
\Ladd{and finally another row vector containing the multi-index from which the tensor grid is generated.}

\Ladd{Note, however, that there is some redundancy in the information stored in the structure array.
  Specifically, the same knot might appear in more than one tensor grid and thus be stored more than once
  (for instance, the knot $(0.5, 0.5)$ appears in all of the tensor grids forming the sparse grids on both rows of Figure \ref{fig:SG_comb});}
this phenomenon is even more pronounced (actually, desired!) when nested sequences of knots are used.
Therefore, it is useful to have at disposal a compact representation of a sparse grid, where duplicate knots are stored only once, together with their ``lumped'' quadrature weights,
obtained taking the linear combination of the quadrature weights of each instance of the repeated knot with the combination coefficient weights in \eqref{eq:sg_interp}.
This representation is called reduced sparse grid \Ladd{and is obtained by detecting} (up to a certain tolerance, tunable by the user) the identical knots,
\Ladd{deleting} the possible repetitions, and \Ladd{computing} the corresponding quadrature weights.
The resulting data structure is a single \Ladd{structure} (see \Ladd{Figure} \ref{fig:sg_data_structure}c), with five fields:
a matrix \Ladd{where each collocation knot of the sparse grid appears only once (each knot being a column)},
a vector for their corresponding ``lumped'' quadrature weights,
an integer containing the number of knots/weights, 
and finally two vectors of indices mapping from the \Ladd{list of knots with repetitions to} the reduced version and vice-versa. 

Note that both extended and reduced formats are useful for working with sparse grids and should always be stored in memory.
This implies a certain redundancy in memory storage, and reducing a sparse grid takes a non-negligible computational time,
as we further discuss in Example \ref{example:generation-cost} below. 
However, having the two structures at hand considerably simplifies coding operations
on sparse grids such as interpolation 
or conversion to Polynomial Chaos Expansion (these operations are described in details in \Ladd{Section} \ref{sect:features}), 
and hides a lot of complexity from the final user. 

\begin{remark}
  Once the family of knots and the level-to-knots function are known, it would be in principle possible to 
  perform the sparse grid reduction comparing indices of knots rather than coordinates, which is faster and does not need to use a tolerance.
  However, we decided to go for the \Ladd{admittedly} slower alternative and compare coordinates rather than indices, \Ladd{as} this
  makes it easier for a generic user to introduce their own family of collocation knots.

  Indeed, especially for non-nested knots, determining if sets of knots at different levels (not just consecutive ones) have knots in common,
  i.e., assessing $\mathcal{T}_{n,k} \cap \mathcal{T}_{n,j}$ for generic choices of $k,j \in \mathbb{N}$ is not straightforward.
  This information \Ladd{would however be} needed if we were to use only indices when reducing a grid,
  thus the user would need to provide such information in a suitable format (and possibly precompute it offline -- again up to a tolerance),
  which might be not easy to do.

  Comparing coordinates instead allows much more flexibility in the way knots are provided by the user.
  In particular, the user does not even need to precompute knots offline, but \Ladd{can} just provide a function that
  computes them at each call, knowing that the reduce operation is robust to \Ladd{differences} in coordinates induced by floating point
  \Ladd{arithmetic}.
  This is actually what happens for Clenshaw--Curtis and Gaussian knots that are not precomputed and tabulated but rather computed at each call.

\end{remark}

\begin{example}[Computational cost]\label{example:generation-cost} 
  \Ladd{Here we} measure memory usage and CPU time
 for generating Smolyak sparse grids (i.e., using \eqref{eq:TD_set_aka_smolyak} to generate the multi-index set and the level-to-knots function in \eqref{eq:m-doubling})
 \Ladd{in reduced format, for increasing $N=2,\ldots,10$ and for fixed $w=3$ or $w=5$.}
 In this example, we use the so-called Clenshaw--Curtis knots, a choice that ensures that the grids generated will be nested.
 We display the \Ladd{growth of the size of the} resulting sparse grid 
 and the computational time in \Ladd{Figure} \ref{fig:sg_cost}a, and the percentage of computational time
 taken by the reduction step in \Ladd{Figure} \ref{fig:sg_cost}b. 

 We then \Ladd{swap the role of $w$ and $N$ and} perform the dual experiment, i.e., we measure memory usage and CPU time for generating reduced Smolyak sparse grids
 for fixed $N=3$ or $N=5$ and increasing $w=2,\ldots,10$. Sparse grid size and total CPU time are now reported in
 \Ladd{Figure} \ref{fig:sg_cost}c while \Ladd{Figure} \ref{fig:sg_cost}d shows the percentage of time taken by the reduction step.

 Both the computational time and the sparse grid size can be seen to grow faster with respect to $w$ than $N$. 
 Moreover, the time taken by the reduction step is \Ladd{only slightly increasing} the total time when keeping $w$
 to small values and increasing $N$ (panel b), whereas steadily increasing with $w$ (panel d).
 This phenomenon is partially due to the chosen level-to-knots function\Ladd{, which doubles the number of points each time $w$ is increased,} and
 using another type of level-to-knots function\Ladd{, which grows points more slowly, could} result in a lower percentage of time being spent on reduction.
	
	\begin{figure}[tbh]
		\centering
		\subfigure[Sparse grid size and total computational time for increasing $N$ (logarithmic scale in the vertical axis)]{\includegraphics[width = 0.33\textwidth]{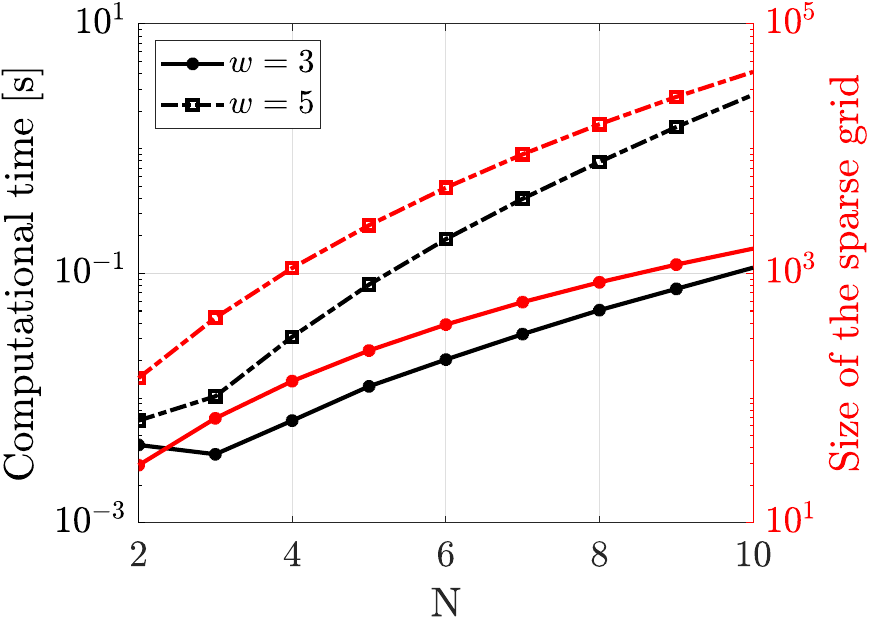}} \hspace{2em} 
		\subfigure[Percentage of computational time taken by the reduction step for increasing $N$]{\includegraphics[width = 0.305\textwidth]{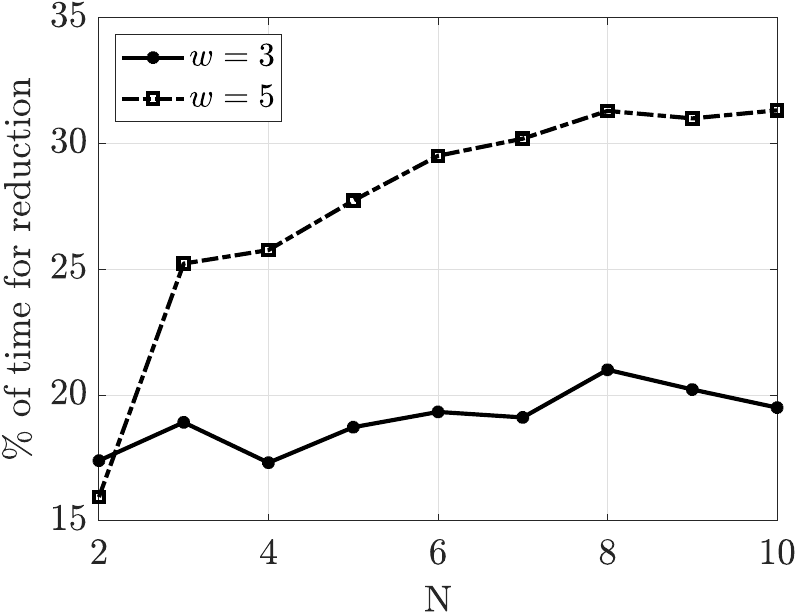}}\\
		\subfigure[Sparse grid size and computational time for increasing $w$ (logarithmic scale in the vertical axis)]{\includegraphics[width = 0.33\textwidth]{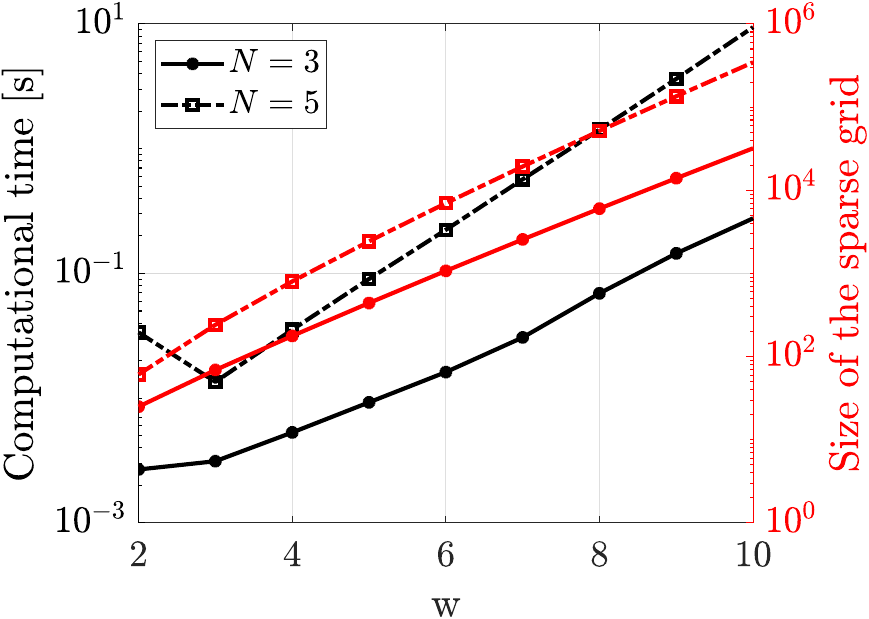}}\hspace{2em} 
		\subfigure[Percentage of computational time taken by the reduction step for increasing $w$]{\includegraphics[width = 0.305\textwidth]{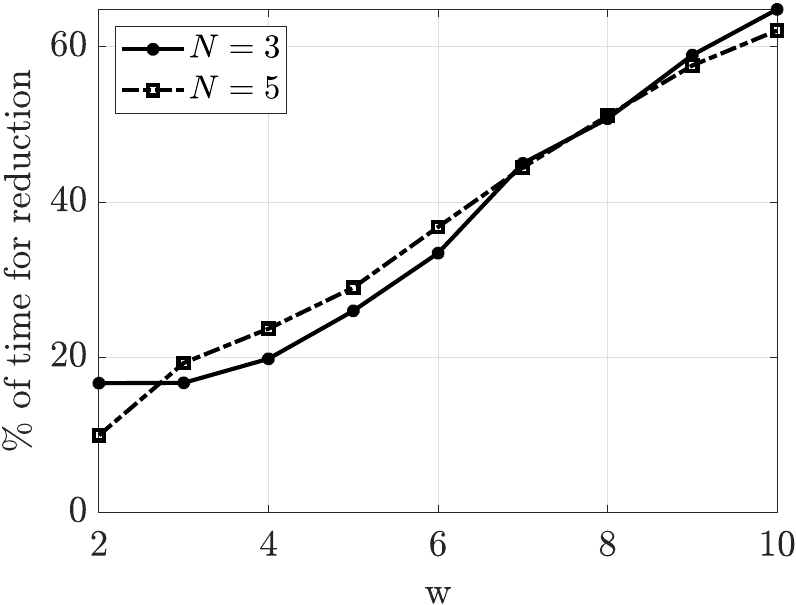}} 
		\caption{Computational cost and size of sparse grids for different values of $N$ and $w$.
                  \Ladd{This test was carried out in Matlab 2019b on a standard laptop with processor Intel(R) Core(TM) i7-8665U CPU 2.10/4.80 GHz and 16 GB RAM.}}
		\label{fig:sg_cost}
	\end{figure}

\end{example}

\subsection{Adaptive sparse grid generation}\label{sect:sparse_grid_generation}

The straightforward way of generating a sparse grid is by specifying \textit{a-priori}
the multi-index set $\mathcal{I}$, i.e., before sampling the function $f$. 
However, as already mentioned, a \textit{greedy adaptive}  approach 
in which the multi-index set (and hence the approximation of $f$) 
is constructed in an iterative way, relying on some heuristic criteria based on the values of the function $f$ obtained so far, is often beneficial.

The adaptive algorithm implemented in the \SGK\ is described in details in \cite{nobile.eal:adaptive-lognormal}
and extends the original by Gerstner and Griebel in \cite{gerstner.griebel:adaptive} \Ladd{in several ways, as listed below};
an alternative approach was proposed by Stoyanov and Webster in \cite{tasmanianPaper1,stoyanov:adaptive2}.

Roughly speaking, the Gerstner--Griebel algorithm starts
with the trivial multi-index set $\mathcal{I} = \{ [1,1,\ldots,1]\}$ and iteratively adds to $\mathcal{I}$ 
the multi-index $\ii$ with the largest heuristic profit indicator
\Ladd{chosen} from a set of candidates, called \Ladd{the} reduced margin of $\mathcal{I}$ and defined as follows:
\[
\mathcal{R}_{\mathcal{I}} = \{\ii \in \mathbb{N}^{N}_+ \text{ s.t. }
\ii \not \in \mathcal{I} \text{ and }
\ii - \ee_n \in \mathcal{I} \; \, \forall \, n \in \{1,\ldots,N\} \text{ s.t. } i_n > 1 \}.
\]
Note that the condition $\ii \in \mathcal{R}_{\mathcal{I}}$ is \Ladd{required} to guarantee 
that $\mathcal{I} \cup \{\ii\}$ is \textit{downward closed}, cf. \eqref{eq:downward_closed}.
The role of the profit indicator is to balance error reduction and additional computational costs brought in by each multi-index $\ii$
(where the cost is measured as the number of new evaluations of $f$ needed to add $\ii$ to $\mathcal{I}$);
in other words, it quantifies the fact that ideally we would like to add to the sparse grid multi-indices that carry a large reduction in
interpolation/quadrature error for a minimal extra cost.
Convergence of this algorithm was recently proved for certain classes of problems, see \cite{eigel.eal:convergence,Feischl2020}.

The computation of the profit of a multi-index $\ii$ actually
requires evaluating $f$ at the new collocation knots that would be added to the sparse grid.
This \Ladd{explains} why the \Ladd{Gerstner--Griebel} algorithm is typically referred to as \Ladd{an} a-posteriori adaptive algorithm,
\Ladd{and it is, in some sense,} a sub-optimal procedure,
since computational work is invested in assessing the profit of multi-indices which might
then turn out to be ``useless''.
A variant of the algorithm, where the computation of the profit 
\Ladd{does} not require evaluating $f$ at the new knots, is proposed in \cite{Guignard:a-post}.
\Ladd{Note however that such profit estimator is tailored to the specific $f$ 
  considered in \cite{Guignard:a-post}. Therefore, this variant of the adaptive algorithm cannot be applied verbatim to any $f$
  but must be suitably adjusted to each case, which is a highly non-trivial task.}

We also mention that \Ladd{the Gerstner--Griebel} algorithm is sometimes referred to as dimension-adaptive: indeed, in the framework introduced so far,
the algorithm will add more knots in the variables that are deemed more important, but these knots are spread throughout the whole support of the random variables
(any clustering being a consequence only of the marginal pdfs $\rho_1,\ldots, \rho_N$) rather than localized in certain regions of the support where
the algorithm has detected local features of $f$.
\Ladd{The latter algorithm would be the locally-adaptive one that we already mentioned in the previous section
  \cite{pflueger:adaptive,Eftekhari:locally.adapt,obersteiner:locally.adapt,ma.zabaras:locally.adapt},
  and that is not supported in the current release of the \SGK.}

\Ladd{As already hinted, the version of the adaptive sparse grid algorithm implemented in the \SGK extends the original by Gerstner and Griebel in several ways, 
  see also Section 3.4.2 of the user manual for a more implementation-oriented discussion and customization options:} 
\begin{itemize}
\item \Ladd{it can use several profit definitions, (see user manual, Table 8)};
\item it can use non-nested knots. \Ladd{This requires introducing suitable profit indicators, 
  as discussed in \cite{nobile.eal:adaptive-lognormal}. Furthermore, note that in this case it will generally happen
  that some of the evaluations of $f$ that have been requested at intermediate iterations will be not used in the final sparse grid approximation,
  because they are associated to multi-indices whose combination coefficient have dropped to zero after having been non-zero for a few iterations.
  In this case however, the evaluations of $f$ on these ``removed'' points are not deleted from memory
  but rather stored in a suitable list of evaluations, from which they will be recovered should
  they be needed in a later iteration (see also the later Section \ref{sect:recycling} about ``evaluation recycling -- recycling from an unstructured list of points'').
  The software allows also to specify whether such search should be done or if the software should just evaluate again $f$
  (which could be faster in case the evaluation of $f$ is very fast and the list of points very large).
  This complexity is however invisible to the end user, who is only in charge of setting correctly the input flags that control the execution
  of the algorithm, namely: to specify that the implementation for non-nested knots must be used,
  and to indicate whether the algorithm should look for knots in the list of ``removed points'' or not
  (see user manual, Table 7).} 
\item it can operate on vector-valued functions, which also requires suitable definitions for the profit indicators, see Equations (12)-(13) of the user manual
    and discussion thereafter;
  \item it \Ladd{implements the the so-called dimension-buffering, see \cite{nobile.eal:adaptive-lognormal} \Ladd{and user manual, Table 7}; this improves the performance of the Gerstner--Griebel algorithm when the function $f$ is ``very high-dimensional'', $N \gg 1$. }
    Indeed, when $N \gg 1$, the size of the reduced margin $\mathcal{R}_{\mathcal{I}}$ grows very quickly,
    which in turn implies a quick growth of the number of evaluations of $f$.
    In this case, if we know that $y_1, \ldots, y_N$ 
    are ``sorted decreasingly according to their importance''
    \Ladd{(this might be, for example, the case when $f$ is the solution of a PDE with uncertain coefficients represented by a Karhunen--Lo\`eve expansion),}
    the \SGK adaptive algorithm starts by exploring only an initial subset of dimensions (the most relevant ones)
      and then gradually adds more dimensions to the approximation, thus limiting the number of indices in the candidate set.
    More specifically, the algorithm splits the random variables in three groups: activated,
    buffered (or non-activated), and neglected.
    A variable $y_k$ is said to be buffered if the algorithm has computed the profit of the
    ``first non-trivial multi-index'' in random variable $y_{k}$, i.e., of $\mathbf{b}_k = [1\,1\,1\,\cdots] + \ee_k$, 
    but $\mathbf{b}_k$ has not been selected yet (i.e., its profit is not the highest one in the list of candidates);
    when $\mathbf{b}_k$ \Ladd{is} selected, \Ladd{it} is moved from the list of candidates
    to $\mathcal{I}$ and the variable $y_k$ becomes activated.
    Then, when the adaptive algorithm is run in ``buffered mode'', with $N_{buf}$ variables:
    \Ladd{
      \begin{itemize}
      \item it begins considering $N_{cur}=N_{buf}$ variables,
        i.e.\ the set of candidate multi-indices is $\mathcal{R}_{\mathcal{I}} = \{ \mathbf{b}_1, \mathbf{b}_2, \ldots, \mathbf{b}_{N_{cur}} \}$
        and all other variables $y_{N_{cur}+1}, y_{N_{cur}+2}, \ldots$ are neglected; 
      \item  as soon as $\mathbf{b}_k$ gets selected (i.e., $y_k$ becomes activated) for some $k \in \{1,\ldots,N_{buf}\}$,
        the number of current variables $N_{cur}$ is increased by $1$ and the first neglected variable $y_{N_{cur}+1}$
        becomes a buffered variable.
    \end{itemize}}
    In this way, at each iteration the algorithm is forced to explore candidates with at most $N_ {buf}$ buffered variables.
    This approach is also discussed in \cite{chkifa:adaptive-interp,schillings.schwab:inverse}, but only for the special case $N_{buf}=1$.
  \end{itemize}

\section{The Sparse Grids Matlab Kit:  operations on sparse grids}\label{sect:features}

 The \SGK implements a number of operations on sparse grids:
  \begin{itemize}
  \item evaluation of $f$ over the collocation knots of the sparse grid;
  \item interpolation (cf. Remark \ref{rem:interpolation}) and quadrature;
    these operations in practice are the solution to the problems of approximating and integrating $f$, cf. beginning of \Ladd{Section} \ref{sect:sg};
    \item computation of gradients and Hessians (by finite differences) of the sparse grid interpolant;
  \item conversion to Polynomial Chaos Expansions (PCE) and computation of PCE-based Sobol indices for sensitivity analysis.
  \end{itemize}
  We discuss below the most noteworthy features implemented in the code, and refer the reader to the \Ladd{Section 4 of the user manual}
  for a thorough discussion of each functionality with practical examples. 

\subsection{Evaluation recycling}\label{sect:recycling}

Evaluation of $f$ on the collocation knots of a sparse grid is of course conceptually straightforward -- it is simply a matter of
looping through the knots of a sparse grid and calling the evaluation of $f$ on each of them. To this end, the \SGK provides
a convenience wrapper function, to which $f$ is passed as anonymous function, \Ladd{see Section 4.1 of the user manual}.
This wrapper can take as input a list of collocation knots
where the function has already been evaluated (either another sparse grid or, \Ladd{say,} coming from a Monte Carlo sampling)
and will detect which of these evaluations can be ``recycled'', to reduce the amount of calls to $f$
\Ladd{(see user manual, Section 4.2.1).}
This algorithm proceeds in two different ways depending on whether the list of knots is another sparse grid or not.
\begin{itemize}
\item If the list of knots is actually another sparse grid (which needs to be passed both in reduced and non-reduced format),
  \Ladd{and both sparse grids are using the same family of univariate collocation knots}: \Ladd{the search for knots in common is speeded up by} comparing essentially the multi-indices in the two grids
  rather than comparing the actual coordinates of the knots
  (i.e., comparing mostly integer numbers, which is of course faster than comparing floating point numbers). 
  Of course, larger savings are obtained if nested collocation knots are used when generating the two grids.
\item If the list of knots is unstructured (\Ladd{say}, a simple list of knots): the same algorithm
  used when reducing a sparse grid is employed to compare the coordinates of the knots in the list with \Ladd{those} in the sparse grid\Ladd{, cf. Section \ref{sect:sg_data_structure}}.
\end{itemize}

\begin{example}[Recycling]
  In \Ladd{Figure} \ref{fig:recycling_test}a we compare the number of evaluations of $f$ over a Smolyak grid with Clenshaw--Curtis (i.e., nested) knots
  for $N=2,4,6$ and increasing values of the level $w$, with and without recycling from the previous grid (i.e., recycling the evaluations at level $w-1$
  to evaluate $f$ over the grid at level $w$). The same information is shown in \Ladd{Figure} \ref{fig:recycling_test}b,
  where we display the percentage of evaluations saved when making use of the recycling functionality, and indeed observe that in this case
  the saving is considerable ($30-50\%$ of the evaluations).
    However, note that this amount depends on the type of multi-index set, the level-to-knots function and the type of knots used (nested/non-nested).
  
  \begin{figure}[tb]
    \centering
    \subfigure[Number of evaluations without recycling (solid lines) and with recycling (dashed lines) (logarithmic scale on the vertical axis)]{\includegraphics[width=0.35\textwidth]{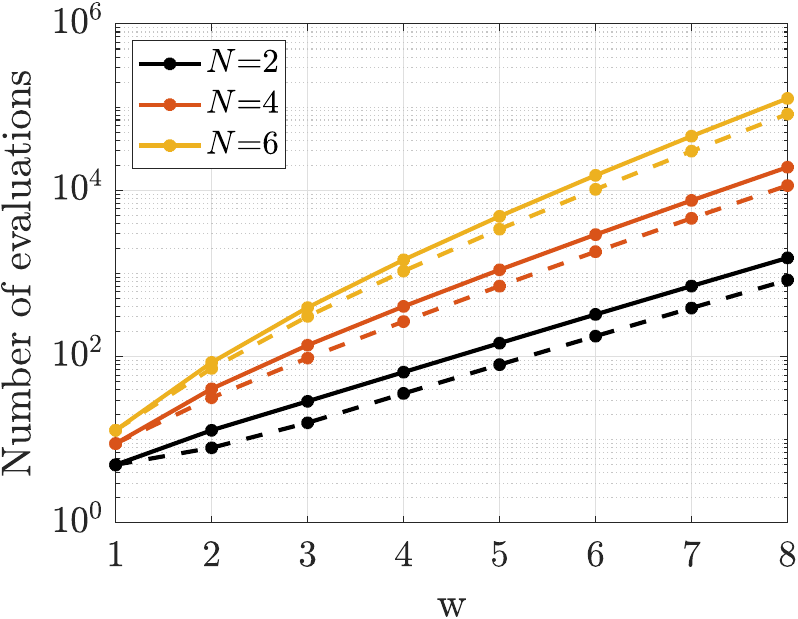}} \hspace{2em}
    \subfigure[Percentage of saved evaluation in case of recycling with respect to the number of evaluations without recycling]{\includegraphics[width=0.35\textwidth]{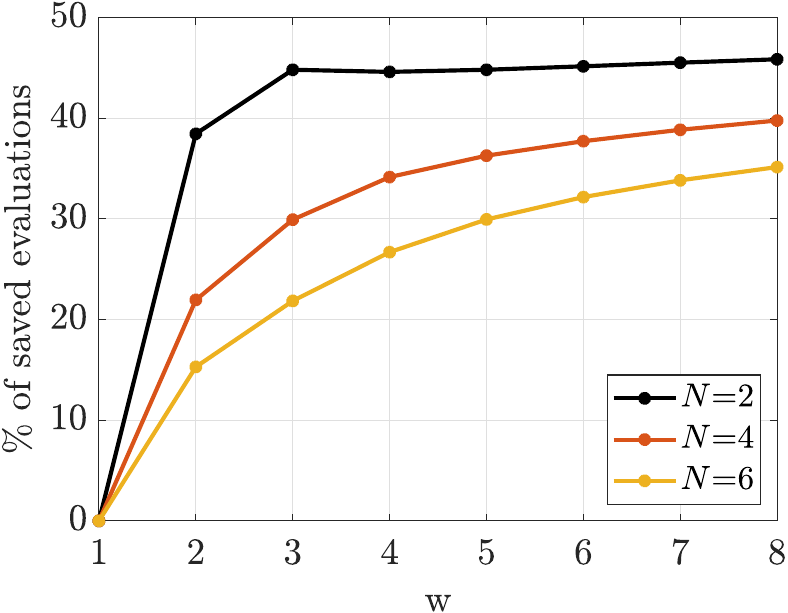}}
    \caption{Assessment of the recycling functionality: test for different values of $w$. }
    \label{fig:recycling_test}
  \end{figure}

\end{example}

\subsection{Parallelization}\label{sect:parallel}

\Ladd{The use of a dedicated wrapper function for evaluating $f$ also allows
  the evaluations of $f$ to be performed in serial or parallel in a way that is transparent to the user (see user manual, Section 4.2.2). 
  As soon as $N_{par}$ or more evaluations are required, the code switches on the Matlab Parallel Toolbox,
  which allocates the requested evaluations of $f$ on all available workers.}
The user controls only the value of $N_{par}$, \Ladd{which} should be set depending on the CPU time required by a single evaluation of $f$.
Indeed, for very fast evaluations of $f$ the communication time between the parallel workers and the central Matlab instance might exceed the evaluation time.
We conclude this paragraph mentioning some technical aspects:
\begin{itemize}
\item The allocation of the evaluations of $f$ on the workers
  is completely delegated to \Ladd{Matlab's} built-in scheduler, and, in particular, this means that we are assuming
  that evaluating $f$ requires the same CPU time for every value of $\yy$ (which is not necessarily the case when $f$ is
  the result of a complex PDE solver).
\item The only part of the \SGK that makes explicit use of the Matlab parallel environment is the evaluation of $f$ over the grid knots.
  All other operations (sparse grid generation and reduction, other operations on sparse grid mentioned at the beginning of this section)
  do not have an explicit parallel implementation.
\item If the adaptive algorithm is used, at each iteration several candidate indices are tested by adding them to the current sparse grid.
  \Ladd{The question that then arises is} whether to consider candidate indices sequentially and then execute
  \Ladd{in parallel} only on the knots requested by each index
  or conversely to gather all new knots requested by all candidates and parallelize them all in a single loop.
  We implement the first strategy, since a) in this way we can use the fast version of the evaluation recycling
  (indeed, the union of all \Ladd{the} new knots needed is not a sparse grid, whereas by testing one index at a time we can compare two sparse grids:
  \Ladd{the one with and the one without such index}) and b) it improves code modularity. 
\end{itemize}

\subsection{Interface with external software by UM-Bridge protocol}\label{sect:external-software} 


\Ladd{A further advantage of using a wrapper function for evaluations of $f$ is that external software
for evaluating $f$ can be simply connected to the \SGK: to this end, it is enough to
encapsulate calls to such external software in an anonymous function, and to
pass the latter to the wrapper routine.}

A particularly efficient way \Ladd{is by using} the UM-Bridge software \cite{Seelinger2023:umbridge},
which implements a standardized HTTP protocol (available in several languages, including Matlab)
\Ladd{for calling external software to evaluate $f$ from within outer loop software, such as UQ and optimization software.
 Using UM-Bridge to call external software from \SGK can take as little as 3 lines of code, see Section 4.5.2 of the user manual.}

\Ladd{Communication} via HTTP messages also allows \Ladd{for the interfacing of}
the \SGK with software running on remote servers. In particular, if \Ladd{these servers allow} parallel requests,
activating \Ladd{parallel execution from within Matlab} as discussed above will automatically enable parallelism on the servers.
More specifically, the \SGK\ \Ladd{would activate one or more Matlab workers, whose only task would be to control a corresponding processor on the remote server,
  to which the actual evaluation of $f$ is delegated.}
Since the workload of the Matlab workers is minimal (just sending an HTTP message)
one can activate \Ladd{a large number of} workers, opening the door to large-scale UQ applications,
\Ladd{see} \cite{seelinger.eal:kubernetes} for a naval engineering UQ application, in which the solver runs remotely in parallel on the Google Cloud Platform. 


\subsection{Interpolation and quadrature}\label{sect:interp+quad}

\Ladd{The \SGK provides an interpolation function that evaluates \eqref{eq:sg_interp}, see Section 4.1 of the user manual for details.} 
The implementation follows closely the mathematical formulation:
\begin{itemize}
\item it loops through the tensor grids, creating a Lagrange interpolant of $f$ on each tensor grid \Ladd{as in \eqref{eq:interp_tensor},}
  and evaluating \Ladd{it} at the requested knots (the standard form of Lagrange 
  polynomials \Ladd{\eqref{eq:lagrange_pol}} is implemented in the \SGK \,-- another possibility would be to implement their barycentric form); 
\item the evaluations on each tensor grid are combined with the combination technique coefficients $c_{\ii}$.
\end{itemize}
Note that for this \Ladd{procedure} both the extended and reduced versions of the sparse grid are needed.
\Ladd{This is due to the fact that the information about tensor grids needed for the above-listed operations
  (knots, combination technique coefficients) is stored in the extended format,
  whereas the values of $f$ on the sparse grid knots are stored in compressed format, i.e.,
  in vectors (matrices for vector-valued $f$) whose number of elements (columns for vector-valued $f$)
  is the number of knots without repetitions in the reduced grid.
  The index information stored in the reduced grid (see Figure \ref{fig:sg_data_structure}c) is therefore needed to be able to assign the correct values of $f$ at each node of each tensor grid. 
}

Similarly, \Ladd{the \SGK also provides an implementation for the quadrature formula in \eqref{eq:sg_quad},
  see again Section 4.1. of the user manual for details.} 
Note however that \Ladd{this} implementation actually does not require looping through the tensor grids,
therefore it does not need \Ladd{both} the extended and reduced format, but only the reduced one.
Indeed, the quadrature weights in the reduced format of the sparse grid already take into account the possible occurrences
of the knots in multiple grids \Ladd{and} the combination technique coefficients.
Therefore the only operation \Ladd{required is a} linear combination of the evaluations of $f$
with the quadrature weights stored in the reduced grid format.

\subsection{Polynomial Chaos Expansion and Sobol indices computation}\label{sec:PCE_and_sobol}

The sparse-grid approximation of a function $f$ is based on Lagrange interpolation polynomials, and hence is a nodal approximation. 
However, sometimes it is interesting to work with modal approximations instead, specifically with the generalized Polynomial Chaos Expansion (gPCE)
\cite{xiu.karniadakis:wiener,ernst:approximability,sudret:sobol}, i.e., an expansion of $f$ over multi-variate $\rho$-orthonormal polynomials 
\cite{gautschi:orthogonal,szego:orthogonal}, \Ladd{that gets typically truncated as}:
\begin{equation}\label{eq:PCE}
f(\yy) \approx \sum_{\pp \in \Lambda} d_{\pp} \mathcal{P}_{\pp}(\yy).
\end{equation}
\Ladd{Here} $\Lambda \subset \Nset^N$ is a multi-index set \Ladd{(note that this time multi-indices can thus have zero entries)}
and $\mathcal{P}_{\pp} = \prod_{n=1}^N P_{p_n}(y_n)$ are products of $N$ univariate $\rho_n$-orthonormal polynomials of degree $p_n$.
The multi-index set $\Lambda$ can be prescribed either a-priori based on the regularity of $f$ \cite{back.nobile.eal:comparison,shen.wang:sparse}
or adaptively \cite{sudret:adaptive.PCE.with.reg,chkifa:adaptive-taylor} \Ladd{and}
the coefficients $d_{\pp}$ can be computed in several ways, \Ladd{for example,} by quadrature \cite{xiu:compute.gPC.coeff.with.sc},
least squares fitting \cite{sudret:adaptive.PCE.with.reg}, or compressed sensing approaches \cite{hampton.doostan:compressed}.

The strategy provided by the \Ladd{\SGK to obtain a gPCE expansion consists of}
computing both the multi-index set $\Lambda$ and the coefficients $d_{\pp}$
by re-expressing the sparse-grid interpolant \Ladd{$\mathcal{U}_{\mathcal{I}} (\yy)$} over the $\rho$-orthogonal basis of choice;
in other words, the \SGK performs a change of basis to represent the same polynomial from a linear combination of Lagrange polynomials
(the sparse-grid interpolant) to a linear combination of $\rho$-orthogonal polynomials \Ladd{(the gPCE)}.
The algorithm that performs the conversion was introduced in \cite{feal:compgeo} (see \cite{constant.eldred.phip:pseudospec} for a similar approach)
and proceeds in two steps:
\begin{itemize}
\item each tensor interpolant $\mathcal{U}_{\ii}$ in the sparse-grid approximation \eqref{eq:sg_interp}, 
  is converted into a linear combination of $N$-variate $\rho$-orthogonal polynomials. \Ladd{This} requires solving
  the following Vandermonde-like linear system for each tensor interpolant:
  \begin{equation}\label{eq:gPCELinearSystem}
    \sum_{\substack{\pp \in \Nset^N :\\ \pp \leq m(\ii) -\mathbf{1} }} \widetilde{d}_{\pp} \mathcal{P}_{\pp} (\yy_k)   = \mathcal{U}_{\ii}(\yy_{k})
    \quad \forall \yy_k \in \mathcal{T}_{\ii};    
  \end{equation}
\item if the same polynomial $\mathcal{P}_{\pp}$ is generated by more than one tensor interpolant,
  the corresponding coefficient $d_{\pp}$ in the final gPCE expansion \eqref{eq:PCE} is the linear combination of the partial coefficients
  $\widetilde{d}_{\pp}$ with coefficients $c_{\ii}$ of the combination technique, see again \eqref{eq:sg_interp}.
\end{itemize}
Also in this case, we need to loop over the tensor grids, therefore both the extended and reduced format of a sparse grid
are needed in the implementation.
Note that the algorithm works in such a way that $\Lambda$ is completely determined by the choices of the level-to-knots function
and of the multi-index set of the sparse grid from which the conversion procedure begins.
The condition number of the Vandermonde-like linear systems \Ladd{\eqref{eq:gPCELinearSystem}} depends on the choice of the families of collocation knots used
to build the tensor grids. In particular, the matrix becomes orthogonal if the tensor grids are
built using $\rho$-Gaussian collocation knots and we want to compute the gPCE expansion over the
the corresponding $\rho$-orthogonal polynomials (\Ladd{for example,} Gauss--Legendre knots and Legendre polynomials,
Gauss--Hermite knots and Hermite polynomials, etc).

The \SGK supports conversion to Legendre, Hermite, Laguerre, generalized Laguerre, and probabilistic Jacobi polynomials,
which are\Ladd{, respectively,} the $\rho$-orthogonal polynomials for uniform, normal, exponential, gamma, and beta probability density functions (all random
variables for which the \SGK provides collocation knots for generation of the corresponding sparse grids), \Ladd{see user manual, Section 4.4.}
Conversion to Chebyshev polynomials is also available, since they are a valid alternative to Legendre polynomials
for expanding functions with respect to the uniform measure, even though they are not orthogonal with respect to it.
The evaluation of the \Ladd{above listed} polynomials is obtained by means of the well-known three-term \Ladd{recurrence} formulas, see \cite{gautschi:orthogonal}.

An example of a situation when having the gPCE expansion of $f$ is helpful
is the computation of the Sobol indices for global sensitivity analysis of $f$ \cite{sobol2001,archer.saltelli.sobol:anova}.
An efficient way to compute such Sobol indices is to perform some algebraic manipulations
on the coefficients of the gPCE, $d_{\pp}$, see \cite{sudret:sobol,feal:compgeo};
to this end, the \SGK provides a wrapper function
\Ladd{(see Section 4.4. of the user manual) in order to perform the required} algebraic manipulations. 
Another reason to perform the conversion to gPCE is to inspect the spectral content of the sparse grid approximation,
to verify how much the nodal representation is storing ``redundant information'', see, \Ladd{for example,} \cite{ernst.eal:collocation-logn}.

\section{Comparison with other software}\label{section:comparison}

\begin{table}[t]
	\centering
	{\setstretch{0.8}
		\begin{tabular}{>{\raggedright}p{2.5cm}>{\raggedright}p{2.3cm}>{\raggedright}p{2.6cm}>{\raggedright}p{2.6cm}>{\raggedright}p{2.3cm}}
			\textbf{Feature} 			& \textbf{Tasmanian} 	& \textbf{SG++} 	& \textbf{spinterp} 	& \textbf{\SGK} 	\tabularnewline \hline \tabularnewline[-5pt]
			\textbf{Native Matlab/Interface}	& interface	 	& interface	 	& native	 	& native	 	\tabularnewline[12pt] 
			\textbf{Currently maintained}	& yes 		 	& yes		 	& no		 	& yes		 	 \tabularnewline[5pt] \hline
			\textbf{Comb.\ tec.\ / hierarchical }	& both 			& mainly hierarchical$^\flat$ 	& hierarchical	& comb.\ tec.	 \tabularnewline[12pt] 
			\textbf{Lagrange basis}		& yes 			& yes			& yes			& yes			\tabularnewline[5pt] 
			\textbf{Piecewise pol. basis}	& yes 			& yes			& yes			& no			\tabularnewline[12pt] 
			\textbf{Splines basis}		& no 			& yes			& no			& no			\tabularnewline[5pt] 
			\textbf{Trigonometric basis}	& yes 			& no			& no			& no			\tabularnewline[12pt] 
			\textbf{Non-nested knots?}		& yes			& no			& no			& yes			 \tabularnewline[5pt] \hline
			\textbf{Dimension-adaptive}		& Webster-Stoyanov 	& Gerstner--Griebel	& Gerstner--Griebel$^\natural$& Gerstner--Griebel \tabularnewline[12pt] 
			\textbf{with non-nested knots}	& no	 		& no		 	& no		 	& yes		 	 \tabularnewline[12pt] 
			\textbf{Buffering}			& implicit$^\dag$	& no			&no 			& yes		 	 \tabularnewline[5pt] 
			\textbf{Local adaptivity}		& yes		 	& yes		 	& no		 	& no	\tabularnewline[5pt] \hline \tabularnewline[-5pt]
			\textbf{Parallel eval. of $f$}	& OpenMP, CUDA/Hip 	& OpenMP	 	& no		 	& Matlab Parallel Toolbox \tabularnewline[12pt] 
			\textbf{knot recycling}		& yes		 	& yes		 	& yes		 	& yes		 	 \tabularnewline[5pt] 
			\textbf{Connection to external $f$} & yes, through LibEnsamble$^\spadesuit$ & no$^\sharp$& no$^\diamondsuit$ 	& yes, through UM-Bridge \tabularnewline[5pt] \hline \tabularnewline[-5pt]
			\textbf{Gradients}			& yes			& yes, in optimization module& yes (exact)	& yes			 \tabularnewline[12pt] 
			\textbf{Hessian}			& no			& no			& no			& yes			 \tabularnewline[5pt] \hline
			\textbf{Random vars. beyond uniform?}& yes			& no$^\sharp$		& no			& yes			 \tabularnewline[12pt] 
			\textbf{Computation of PCE}		& no 			& no$^\sharp$		& no 			& yes		 	 \tabularnewline[12pt] 
			\textbf{Computation of Sobol idx}	& no		 	& yes		 	& no		 	& yes		 	 \tabularnewline[12pt] 
	\end{tabular}}
	\caption{Comparative table of Matlab software for sparse grids and sparse-grids-based uncertainty quantification. Annotations: \\
		$^\flat:$ limited support for \Ladd{combination technique} provided by the \Ladd{combination technique} module \\
		$^\natural:$ with blending of dimension-adaptive and Smolyak construction, and dropping of multi-indices with small profit \cite{spinterp:manual} \\
		$^\dag:$ once the anisotropy estimate in the Webster--Stoyanov algorithm is reliable \\
		$^\spadesuit:$ \url{https://libensemble.readthedocs.io} \\
		$^\sharp:$ supported through interface with Dakota \cite{dakota}, but not in the Matlab interface \\
		$^\diamondsuit:$ only through Matlab system calls }
	\label{tab:comparison}
\end{table}

In this section, we provide a comparison of \Ladd{the \SGK with the packages} for sparse grids and sparse-grids-based UQ
\Ladd{in Table \ref{tab:software} that} either \Ladd{are} natively written in Matlab or provide an interface to Matlab.
We thus compare the \SGK with SG++, Tasmanian and Spinterp; we \Ladd{choose to neglect} UQLab, since \Ladd{it} does not provide 
full sparse grids functionalities (more precisely, it provides sparse grids quadrature but not sparse grids interpolation).
We focus on a comparison in terms of functionalities rather than on computational efficiency
since the typical CPU-intensive utilization scenario of this kind of software is the construction of surrogate models
for UQ purposes, in which case the computational cost is largely dominated by the evaluation of the function $f$,
and only a small fraction of \Ladd{cost may be ascribed} to the actual sparse grid functionalities.


\Ladd{The results of the comparison are reported in Table \ref{tab:comparison}, which} is divided \Ladd{into} several ``thematic'' blocks.
We begin by providing the essential software information: native Matlab/interface and whether \Ladd{each package} is currently maintained or not.
We then move to comparing the basic features of sparse grids provided by each \Ladd{package}, as discussed in \Ladd{Section} \ref{sect:sg}:
the sparse grid forms implemented (combination technique / hierarchical),
the supported basis functions and knots and, more specifically, whether non-nested knots can be used.
The third block focuses on adaptive algorithms: \Ladd{what dimension-adaptive is provided, whether locally-adaptivity is implemented
  (cf. Sections \ref{sect:sg} and \ref{sect:sparse_grid_generation})},
and some connected features, most notably the dimension buffering discussed in \Ladd{Section} \ref{sect:sparse_grid_generation}.
Next, \Ladd{we consider} features connected to evaluations of $f$: evaluation recycling
(\Ladd{Section} \ref{sect:recycling}), parallel evaluation (\Ladd{Section} \ref{sect:parallel}),
and \Ladd{the possibility of connecting} to external software to evaluate $f$ (cf. \Ladd{Section} \ref{sect:external-software}).
The last two blocks deal with additional features: derivatives (i.e., gradients and Hessians) and UQ functionalities.

The main take-away point of this comparison is that the four \Ladd{packages considered have very} little overlap:
they all provide Lagrange-based interpolation and gradient computation, and support recycling evaluation
(which in a way can be considered the minimum \Ladd{requirements of a piece of software that wants to be of any practical use})
and dimension-adpativity. Other than this, they all come with their
own sets on unique features. In general, Tasmanian and the \SGK have more UQ functionalities, whereas SG++ can be thought \Ladd{of} as a more generalistic
sparse grids software, since it provides additional modules that implement functionalities for PDE solving and data mining.
Spinterp provides an interesting implementation of the dimension-adaptive algorithm, that a) blends the Gerstner--Griebel and the Smolyak a-priori
grid construction by setting a so-called balancing parameter that can range from 1 (100\% of sparse grid knot generated by the adaptive algorithm)
to 0 (Smolyak construction) \Ladd{and} b) can drop multi-indices added to the approximation with little profit, see \cite{spinterp:manual}.
Tasmanian provides the largest choice of basis functions, whereas \SGK has some unique functionalities for UQ:
dimension buffering for adaptivity, \Ladd{support for} PCE and Sobol indices, as well as computation of Hessians which could be useful,
\Ladd{for example, for Bayesian inversion tasks (see, for example,
  \cite{piazzola.eal:SIR,thanh-bui.gattas:MCMC,petra.ghattas:ice.sheet.MCMC,stuart:acta.bayesian,kim.eal:hippylib} for more details).}

\section{Conclusions}\label{section:conclusions}
In this manuscript we have introduced the combination technique form of the sparse grid methodology
for approximating and computing integrals of high-dimensional functions, in particular for UQ purposes.
The \SGK is a Matlab \Ladd{package} that can be used to these ends.
We have discussed the data structure of the software and the mathematical
aspects of the functionalities implemented in it, and we have compared it with other Matlab software for
sparse grids and UQ (Spinterp, Tasmanian, SG++). Compared to alternative software,
the \SGK\ \Ladd{provides the most tools for UQ, for example} dimension-buffering for adaptivity, \Ladd{and} support for PCE and Sobol indices.

\section*{Ackowledgments}
  Lorenzo Tamellini and Chiara Piazzola have been supported by the PRIN 2017 project 201752HKH8
  ``Numerical Analysis for Full and Reduced Order Methods for the efficient and accurate solution of complex systems governed by Partial Differential Equations (NA-FROM-PDEs)''.
  Lorenzo Tamellini has been also supported by the Research program CN00000013
  ``National Centre for HPC, Big Data and Quantum Computing -- Spoke 6 - Multiscale Modelling \& Engineering Applications''.
  Chiara Piazzola has been also supported by the Alexander von Humboldt Foundation. 
  The authors gratefully acknowledge several \Ladd{people} who contributed to the
  development of the package either by providing implementations for some functions or by using the software and reporting success cases, bugs and missing features.
  In particular \Ladd{we would like to thank}: Fabio Nobile (early version of the code and continued support throughout the development of the project),
  Alessandra Sordi and Maria Luisa Viticchi\`e (early contributions to the code),
  Francesco Tesei and Diane Guignard (adaptive sparse grids),
  Giovanni Porta (Sobol indices and conversion to PCE), Bj\"orn Sprungk (adaptive sparse grids and weighted Leja knots), Francesca Bonizzoni (compatibility with Octave).
  Finally, we thank Miroslav Stoyanov and Dirk Pfl\"uger for the help in crafting \Ladd{Table} \ref{tab:comparison}.

\bibliography{biblio,numques_biblio}

\end{document}